\documentclass[preprint]{aastex}

\usepackage{natbib}
\usepackage{amsmath}
\usepackage{multirow}
\usepackage{color}

\newcommand{\bLam}{\mbox{\boldmath{$\Lambda$}}}
\newcommand{\bphi}{\mbox{\boldmath{$\phi$}}}
\newcommand{\bpsi}{\mbox{\boldmath{$\psi$}}}
\newcommand{\btheta}{\mbox{\boldmath{$\theta$}}}
\newcommand{\bnu}{\mbox{\boldmath{$\nu$}}}
\newcommand{\bmu}{\mbox{\boldmath{$\mu$}}}

\newcommand{\by}{\mbox{\boldmath{$y$}}}

\newcommand{\bP}{\mbox{\boldmath{$P$}}}
\newcommand{\bA}{\mbox{\boldmath{$A$}}}

\newcommand{\obs}{_{\text{obs}}}

\renewcommand{\r}{\right}
\renewcommand{\l}{\left}
\newcommand{\ind}{\buildrel{\rm indep}\over\sim}
\newcommand{\like}{{\cal L}}

\newcommand{\dirpar}{\psi}
\newcommand{\bdirpar}{\bpsi}

\begin{document}

\title{Detecting Unspecified Structure in Low-Count Images}

\author{
Nathan M. Stein\altaffilmark{1},
David A.\ van Dyk\altaffilmark{2},
Vinay L.\ Kashyap\altaffilmark{3}, 
and Aneta Siemiginowska\altaffilmark{3}
}
\affil{$^1$ Department of Statistics, The Wharton School, University of Pennsylvania, \\
400 Jon M. Huntsman Hall, 3730 Walnut Street, Philadelphia, PA 19104-6340, USA
\email{
natstein@wharton.upenn.edu
}
}
\affil{$^2$Statistics Section, Imperial College London \\
Huxley Building, South Kensington Campus, London SW7 2AZ, UK\\ 
\email{
dvandyk@imperial.ac.uk
}
}
\affil{$^3$Smithsonian Astrophysical Observatory, \\
60 Garden Street, Cambridge, MA 02138, USA
\email{
vkashyap@cfa.harvard.edu \\
asiemiginowska@cfa.harvard.edu 
}
}

\begin{abstract}
Unexpected structure in images of astronomical sources often presents itself upon visual inspection of the image, but such apparent structure may either correspond to true features in the source or be due to noise in the data. This paper presents a method for testing whether inferred structure in an image with Poisson noise represents a significant departure from a baseline (null) model of the image. To infer image structure, we conduct a Bayesian analysis of a full model that uses a multiscale component to allow flexible departures from the posited null model. As a test statistic, we use a tail probability of the posterior distribution under the full model. This choice of test statistic allows us to estimate a computationally efficient upper bound on a p-value that enables us to draw strong conclusions even when there are limited computational resources that can be devoted to simulations under the null model. We demonstrate the statistical performance of our method on simulated images. Applying our method to an X-ray image of the quasar 0730+257, we find significant evidence against the null model of a single point source and uniform background, lending support to the claim of an X-ray jet.
\end{abstract}

\keywords{galaxies: jets---methods: data analysis---methods: statistical---techniques: image processing---quasars: individual (0730+257)---X-rays: general}

\section{Introduction}

Detecting scientifically meaningful structure in digital images is a ubiquitous and notoriously difficult problem. Typically, image analysis algorithms in high-energy astronomy are optimized for the detection and characterization of point sources. However, this strategy fails when confronted with complex extended structures at many scales. 
Optical observations often reveal rich and irregularly structured emission associated with a variety of objects in the universe, such as galaxies, nebulae, clusters of stars, or clusters of galaxies. The X-ray emission of these objects is often as rich as the optical, but the Poisson nature of the observed images makes the emission hard to discern. The X-ray images are often sparse and may require binning to expose the emission features, but binning lowers the resolution and potentially leads to loss of the smaller scale structures. Detecting irregular X-ray emission is thus challenging, and there has been no principled method to date to assess the statistical significance of 
arbitrary irregular features
in X-ray images.

Source detection algorithms, such as {\tt celldetect} \citep{cald:etal:01} and {\tt wavdetect}  \citep{free:etal:02} in CIAO (Chandra Interactive Analysis of Observations), work quite well for detecting point sources, but not for unspecified irregular emission.  The CIAO {\tt vtpdetect} algorithm \citep{ebel:wied:93} can identify extended regions by looking at the distribution of tesselation areas and imposing a threshold cut, but does not otherwise determine the significance of the detected sources. Moreover, {\tt vtpdetect} can spuriously combine the diffuse emission with embedded point sources, resulting in the confusion of the emission components.  Other techniques used by astronomers include direct two dimensional fitting of image features with pre-defined models, and qualitative analysis of residuals from such fits.  Many studies also rely on maximum entropy-based image deconvolution techniques \citep[e.g.,][]{rich:72,lucy:74}, but these typically do not yield unique fits and do not provide associated uncertainties.  A Bayesian method that constructs a representation of an image using a basis set of regions \citep[{\tt pixons};][]{pina:puet:92} has also been tried on astronomical images, but again, without a means to evaluate the significance of identified regions. More generally, powerful computational tools such as Markov chain Monte Carlo (MCMC) enable researchers to fit more and more sophisticated models that capture the complexities of astronomical sources and instruments. On its own, however, MCMC is better suited for fitting a model than for choosing between models or for detection problems \citep[see, however,][]{weinberg:12}. Thus new tools are needed for quantifying statistical significance or goodness of fit in the context of complicated MCMC-fitted models.

This paper addresses the problem of detecting image structure not adequately accounted for by a baseline model that represents features known to be present in the image; from a statistical perspective the baseline model serves as the null hypothesis. This formulation is useful in a wide range of applications. Here and in a forthcoming companion paper \citep{mcke:etal:15}, we consider the problem of detecting X-ray jets emitted from quasars. The baseline model includes only the quasar and a flat background, with no additional emission representing the jet. Because it is difficult to specify parametric models that adequately capture the range of possible appearances of X-ray jets, we use a multiscale model that allows flexible, nonparametric departures from the baseline. Another possible application is detecting dynamic behavior, such as the time evolution of a supernova remnant. In these cases, the baseline model could be constructed using an earlier image, and the goal would be to test whether a later image represented a significant departure from the earlier one. Such applications extend beyond astronomy. Detecting changes in images over time is important in fields ranging from medical imaging to surveillance; see \citet{radke:etal:05} for a review. Finally, we might be interested in detecting fine structure blurred by a point spread function (PSF), such as when analyzing filament structure in coronal loops in images of the Sun \citep{mcke:poster}. In this case, the baseline model could include readily apparent low-frequency structure, and the goal would be to detect mid-frequency departures from this model, after removing high-frequency noise.

Much previous work has attempted to quantify the statistical uncertainty of inferred features in noisy images. In functional magnetic resonance imaging, for example, \citet{friston:etal:95} proposed ``statistical parametric maps,'' pixel-wise significance tests with subsequent adjustments for multiplicity based on Gaussian random fields. In astronomy, \citet{ripley:sutherland:90} used spatial stochastic processes to directly model structures in spiral galaxies, and \citet{esch:etal:04} obtained uncertainty estimates for reconstructed X-ray images using a Bayesian model known as EMC2 that included multiple levels for instrumental effects, background contamination, and a multiscale hierarchical representation of the source. \citet{sutton:wand:06} and \citet{sutter:etal:14} used Bayesian models to perform image reconstruction with uncertainty estimates using radio interferometry data. Bayesian methods have also been employed to quantify the uncertainty in the large-scale structure of the Universe \citep{jasche:wand:13} and in secondary anisotropies of the cosmic microwave background radiation \citep{bull:etal:14}.
\citet{friedenberg:genovese:13} proposed a multiple testing procedure for detecting sources in astronomical images. Other approaches can be found in the computer vision literature; see for instance \citet{godtliebsen:etal:04}, \citet{holmstrom:pasanen:12}, and \citet{thon:etal:12}. 

Rather than estimating the uncertainty in inferred image features, we focus on the more fundamental problem of feature detection. Specifically, we adopt a hypothesis testing framework to address the statistical question of whether there is sufficient evidence to conclude that the baseline (null) model is unlikely to have produced  by chance the structure observed in an image. This framework ensures that we can control the probability of a false positive result, i.e., declaring that there is significant additional structure in the image beyond the baseline model, when in fact there is none. Our test statistic is a tail probability of a Bayesian posterior distribution under a full statistical model that includes both the baseline model and a multiscale component for structure not included in the baseline. This  distinguishes our method from existing goodness-of-fit tests for inhomogeneous (baseline) Poisson processes \citep[e.g.,][]{guan:08}. We do not frame our approach in terms of Bayesian model selection (e.g., using Bayes factors) because the flexible full model for additional emission beyond the baseline is intentionally weakly specified, making it especially difficult to reliably apply Bayes factors due to their sensitivity to prior distributions \citep[e.g.,][]{vand:12scma}.

Our framework provides a reference distribution with which we can quantify how inferences given the observed data differ from inferences given data generated under the null model, when all analyses are performed under the full model. Computationally, this is accomplished by simulating multiple replicate images under the null model, fitting the full model to each, and computing the test statistics for each. This gives us a reference distribution for the test statistic that we can compare with the test statistic computed from the observed image in order to determine the statistical significance of apparent structure in the image. A primary novelty of our method is its use of an upper bound on the p-value that enables us to obtain statistical significance with a limited number of replicate images. Because each replicate image is fit under the fully Bayesian model, limiting their number is important for controlling the computational demands of the method.

We use a Bayesian model to infer image structure because it provides a principled way to account for a PSF, varying probability of photon detection, and varying amounts of smoothing at different image resolutions.
This model builds on \citeauthor{esch:etal:04}'s EMC2 in that it uses the same multiscale representation but for a different purpose. Whereas \citet{esch:etal:04} used this multiscale model to fully represent the source, we include a baseline model for the source and use the multiscale model of EMC2 to flexibly capture added structure beyond the baseline model. 
Combining this extension with the formal statistical testing proposed here 
leads to a new statistical software package called LIRA\footnote{LIRA is a package for the R statistical programming language ({\tt r-project.org}) and is available at {\tt github.com/astrostat/LIRA}.} (Low-count Image Reconstruction and Analysis). Like EMC2, LIRA deploys Markov chain Monte Carlo for fully Bayesian model fitting of the overall multilevel statistical model. 

This article is organized into five sections. We begin in Section~\ref{sub:example} with a simulated example that illustrates the scientific problem we aim to solve and serves as a running example for the remainder of the paper.  In Section~\ref{sec:model} we formulate the baseline and full models that we compare using a formal hypothesis testing framework in Section~\ref{sec:testing}.  A set of simulation studies and an analysis of the X-ray jet associated with the 0730+257 quasar illustrate our proposed methods in Section~\ref{sec:numerical}. We conclude with discussion in Section~\ref{sec:disc} and technical details in Appendix~A.

\subsection{A simulated example}\label{sub:example}

As a concrete example of the scientific problems this paper addresses, we consider jet detection in an X-ray image of a quasar. This example applies also to detecting any secondary faint source in the vicinity of a bright point source, such as multiple sources, halos, non-uniform faint emission, and so on.  Quasar jets extend out to distances on the order of 100 kiloparsecs from a supermassive black hole, and can trace the history of the black hole's activity and power \citep{urry:95,harris:06}. Many such jets have been observed for the first time in X-rays with the {\it Chandra} X-ray Observatory,
where some quasar images show a bright point source and a much fainter jet feature.  The jet surface brightness is non-uniform, so brighter knots and fainter emission are often seen along the jet.

\begin{figure}[t]
% \vspace{-.4cm}
\centering

\begin{minipage}[b]{0.4\linewidth}
\centering
(a)

\includegraphics[width=\textwidth]{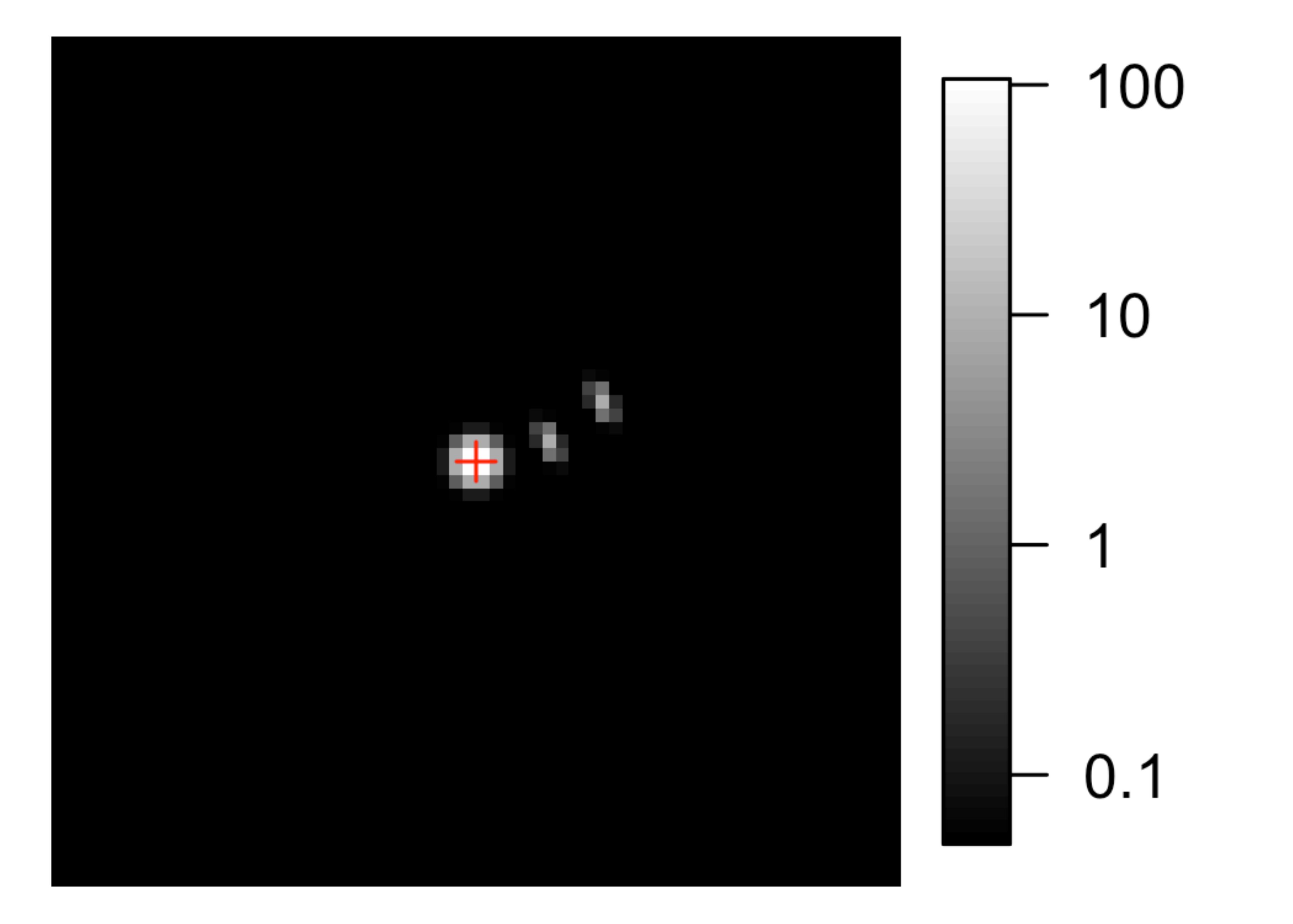}
\end{minipage}
\hspace{.05\linewidth}
\begin{minipage}[b]{0.4\linewidth}
\centering
(b)

\includegraphics[width=\textwidth]{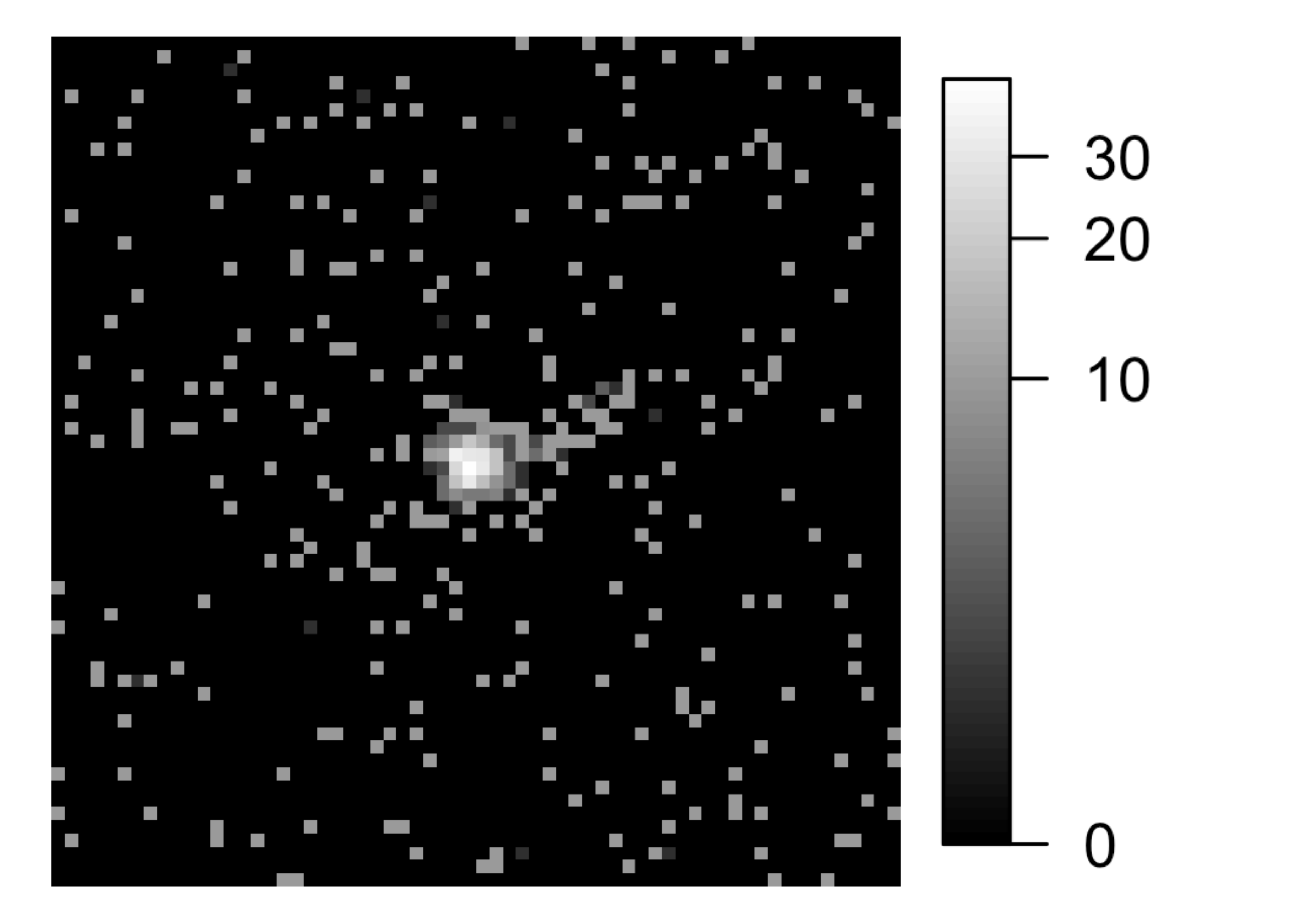}
\end{minipage}

\vspace{.3cm}

\begin{minipage}[b]{0.4\linewidth}
\centering
(c)

\includegraphics[width=\textwidth]{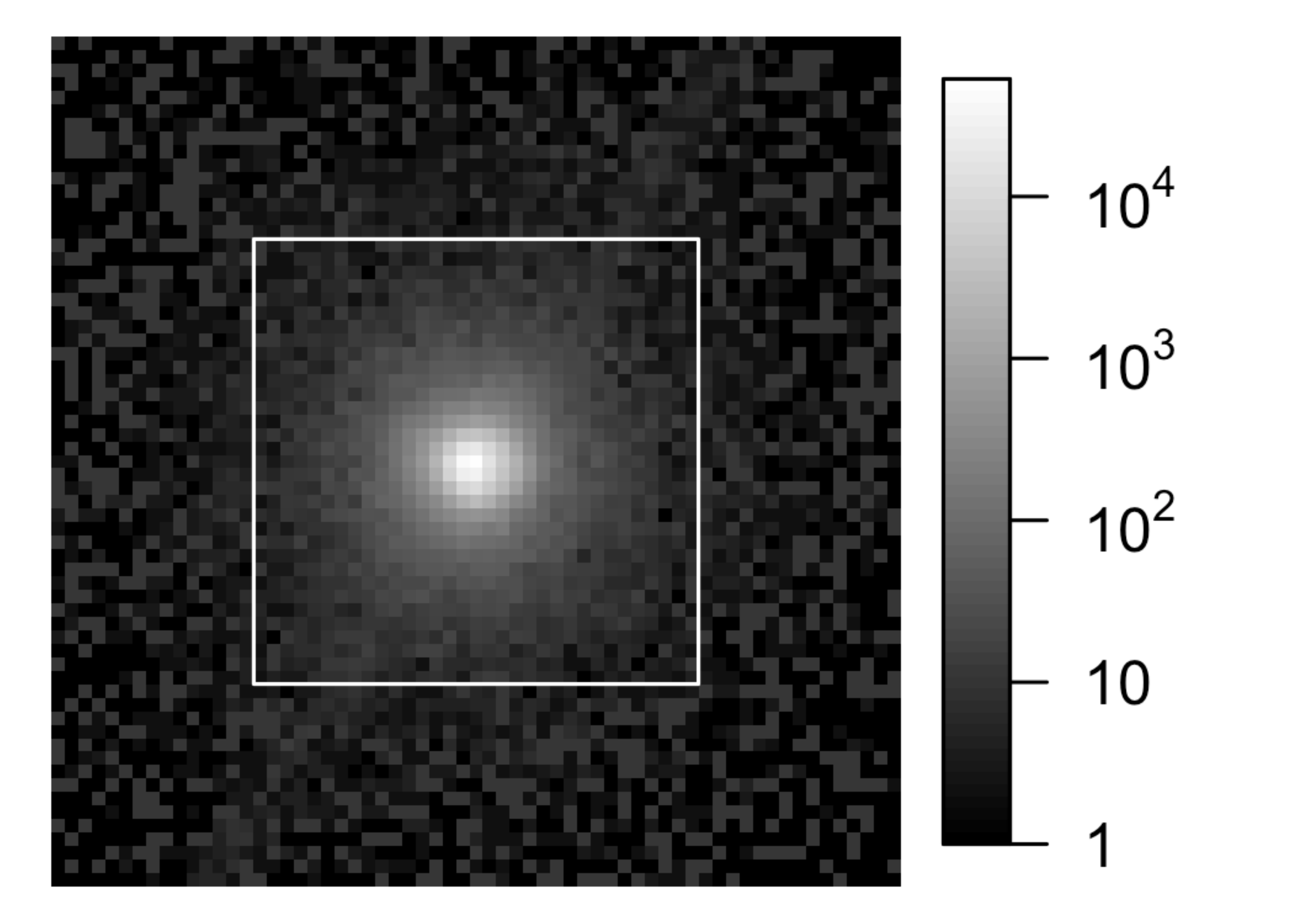}
\end{minipage}
\hspace{.05\linewidth}
\begin{minipage}[b]{0.4\linewidth}
\centering
(d)

\includegraphics[width=\textwidth]{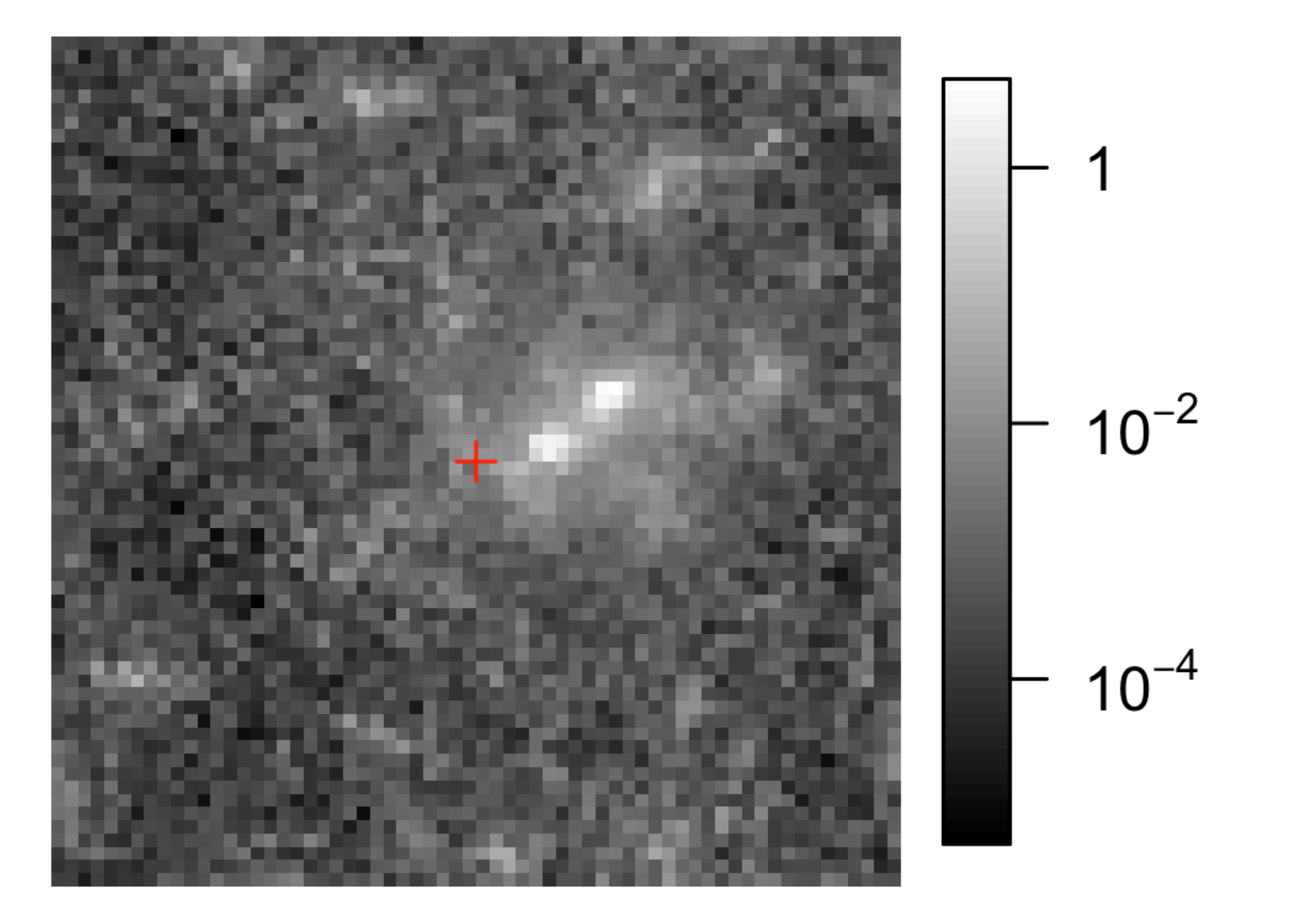}
\end{minipage}

\caption{(a) True underlying intensity for a simulated quasar with a jet consisting of two additional Gaussian sources.
 (b) Simulated observation for the quasar and jet in (a), after convolving (a) with the PSF in (c) and introducing Poisson noise. 
 (c) Un-normalized PSF for the analysis of simulated and observed quasars. The full PSF was used to simulate data, but only the portion inside the white square was used in analyses under the full model.  (d) Pixel-wise posterior means of the added component $\tau_1 \bLam_1$ in the full model, given the simulated data in (a). The red + in (d) identifies the location 
 of the simulated quasar in (a).}
\label{fig:images}
\end{figure}

Figure~\ref{fig:images}(a) is a simulated ground-truth image of a quasar, modeled as a single bright point source with a jet composed of two fainter point sources. Figure~\ref{fig:images}(b) is a simulated observation designed to mimic the degradation of Figure~\ref{fig:images}(a) due to a detector's PSF and Poisson noise from a limited exposure time. Both images are $64 \times 64$ pixels. The quasar was simulated as a two-dimensional spherical Gaussian with standard deviation 0.5 pixel and 500 expected counts. The jet was composed of two elliptical Gaussian components each with ellipticity 0.5, standard deviation 0.5 pixel, and 20 expected counts. The simulated background was uniform across the image with 200 total expected counts (approximately 0.05 counts per pixel).

We aim to compare a baseline model that posits that Figure~\ref{fig:images}(b) was generated from a background-contaminated single point source (i.e., just a quasar, with no jet) with a full model that allows for unspecified structure in the image beyond the single point source. Further, we aim to quantify the statistical evidence in the image for choosing between the two models. Our method specifically avoids parametric modeling of the extended structure (here the jet) because in practice we often do not wish to specify the precise nature of possible departures from the simple model. More generally, we want  to flexibly detect and quantify the evidence for departures from simple baseline models of images that are observed under the imperfect conditions that often arise in high-energy astrophysics, such as photon scatter, background contamination, and non-negligible Poisson noise.

\section{Models and hypotheses}\label{sec:model}

\subsection{The statistical model}\label{sub:full}

We consider an image composed of $n$ photon counts arranged into a grid of pixels; we denote the counts in the $n$ pixels by $\by\obs = (y_1, \ldots, y_n)$. If the two-dimensional image written in matrix form has $l$ rows and $m$ columns, then in our vectorized notation it has dimension $n=lm$. We model the image as a superposition of two Poisson processes. The first is intended to represent known or presumed aspects of the image, which could include anything from background noise to complicated structures of interest. For example, if we aim to quantify the evidence for a jet extending from a known point source, as in Section~\ref{sub:example} and Figure~\ref{fig:images}(b), then the first Poisson process would consist of the point source and background contamination. We refer to this first Poisson process as the \emph{baseline component}. The second Poisson process is intended to account for image features unexplained by the first process and is called the {\it added component}. In the example of testing for a jet, the added component would model the hypothesized jet.

Because of blurring and varying instrument response across the detector, the distribution of the counts observed in detector pixel $i$ is
\begin{equation}\label{eq:detected}
  y_i \ind  \text{Poisson}\l(\sum_{j=1}^n P_{ij}A_j\l(\mu_{0j} + \mu_{1j}\r)\r),
\end{equation}
where $\bP$ is the $n \times n$ PSF, with $(i,j)$ element $P_{ij}$ denoting the probability that a photon in location $j$ is observed in pixel $i$; $\bA = (A_1, \dots, A_n)$ is the detector efficiency, with $A_j$ equal to the probability that a photon in pixel $j$ is detected; and $\bmu_0 = (\mu_{01},\ldots,\mu_{0n})$ and $\bmu_1 = (\mu_{11},\ldots,\mu_{1n})$ are, respectively, the intensities of the baseline and added components. 
The representation of the PSF in Equation~\ref{eq:detected} is quite general in that each column of $\bP$ is the vectorized PSF for a particular source pixel. Thus, this representation allows the PSF to vary across the source. 
Throughout this paper, we assume that $\bP$ and $\bA$ are known, that $\bmu_1$ is an unknown parameter of interest, and that $\bmu_0$ includes some known structure but is partially unknown. When creating the simulated observation in Figure~\ref{fig:images}(b), we set $\bA = (1,\ldots,1)$ and used the PSF in Figure~\ref{fig:images}(c). Details on this PSF appear in Section~\ref{sub:three-images}.

We parameterize component $i$ ($i=0$ or $1$) as $\bmu_i = \tau_i \bLam_i$, where $\tau_i = \sum_{j=1}^n \mu_{ij}$ is the expected photon count in component $i$, and $\bLam_i = (\Lambda_{i1}, \ldots, \Lambda_{in}) = \bmu_i / \tau_i$ is the proportion of $\tau_i$ that is expected in each pixel. The baseline component is often parameterized in terms of a lower dimensional parameter vector $\bnu$, in which cases we write $\bLam_0 = \bLam_0(\bnu)$. The parameter $\bnu$ may include unknown aspects of posited image structure, such as the location of a point source. In practice, some parameters in $\bnu$ may be well-constrained by the data, and fixing these parameters at their estimates causes no problems, but we also consider the situation in which there is substantial uncertainty in at least some components of $\bnu$; see Section~\ref{sub:nuisance}. If the baseline component is fully specified except for its total intensity $\tau_0$, then $\bnu$ will be empty.

In fitting the model in Equation~\ref{eq:detected}, $\tau_1$ and $\bLam_1$ describe the added component and are of direct scientific interest.  The parameter $\tau_0$ is intertwined with these parameters in that the image's total count constrains its total intensity, $\tau_0 +\tau_1$, and thus the fitted $\tau_0$ will decrease as the fitted $\tau_1$ increases. We denote the unknown parameters $\btheta = (\btheta_0, \btheta_1)$, where $\btheta_0 = (\tau_0, \bnu)$ are the parameters of the baseline component and $\btheta_1 = (\tau_1, \bLam_1)$ are the parameters of the added component. Typically, $\bnu$ is a nuisance parameter, at least in the context of searching for added structure beyond the baseline component.

\subsection{Bayesian inference}\label{sub:bayes}

We adopt a Bayesian framework to fit the image parameters, $\btheta$, given the observed photon counts, $\by\obs$. In particular, we quantify our state of knowledge before having seen the data using a {\it prior distribution} and that after having seen the data using a {\it posterior distribution}. Bayes' Theorem allows us to transform the prior distribution into the posterior distribution by conditioning on the observed counts.  In particular, the theorem states that the posterior distribution of $\btheta$ given $\by\obs$ is
\begin{equation}
\pi(\btheta \mid \by\obs)=\dfrac{\like(\by\obs \mid \btheta) \  \pi(\btheta)}{\pi(\by\obs)},
\label{eq:bayes}
\end{equation}
where $\pi(\btheta)$ is the joint prior distribution of $\btheta$, $\like(\by\obs \mid \btheta)$ is the likelihood function of $\by\obs$ given $\btheta$, and $\pi(\by\obs)=\int \like(\by\obs \mid \btheta) \pi(\btheta) {\rm d}\btheta$ is the normalizing constant that ensures that $\pi(\btheta \mid \by\obs)$ integrates to one.

While prior distributions can be used to incorporate external information about the likely values of model parameters, they can also be used to enforce relationships among parameters. We use the added component, for example, to represent structure in an image that does not appear in the baseline component. If we did not impose any constraint on $\bLam_1$, random fluctuation from the baseline component and unstructured background would be indistinguishable from genuine image structures that are missing in the baseline. As detailed in Section~\ref{sub:prior}, we use the prior distribution of $\bLam_1$ to specify a multiscale smooth structure that characterizes the added components that can be identified by our procedure.  

Assuming the prior distributions for $(\tau_0, \btheta_1)$ and $\bnu$ are independent, we can write the posterior distribution as
\begin{equation}
\pi(\btheta \mid \by\obs) \propto \like(\by\obs \mid \btheta) \ \pi(\tau_0, \btheta_1) \ \pi(\bnu),
\label{eq:newbay}
\end{equation}
where we have omitted the denominator of Equation \ref{eq:bayes} because it is a constant determined by the numerator. Since the likelihood function is specified by Equation~\ref{eq:detected}, we need only set $\pi(\tau_0, \btheta_1)$ and $\pi(\bnu)$. Insofar as the baseline model is well specified, the uncertainty in $\bnu$ and  hence the sensitivity of the final result to the choice of $\pi(\bnu)$ are both limited. 
For example, when testing for an X-ray jet in an image of a quasar, $\bnu$ could consist of the location and amplitude of the quasar point source and the intensity of a constant background. These parameters are well constrained by the data and thus relatively insensitive to the choice of prior $\pi(\bnu)$. In practice, we typically use the default settings in CIAO's Sherpa software \citep{freeman:etal:01}
and fit $\bnu$ via maximum likelihood (equivalent to the posterior mode under a uniform $\pi(\bnu)$); see Section~2.4.
We do not further discuss the choice of $\pi(\bnu)$ in this paper, but the choice of $\pi(\tau_0, \btheta_1)$ is central to the general problem and is the topic of Section~\ref{sub:prior}.

\subsection{The prior distributions}\label{sub:prior}

\begin{figure}[t]
\centering
\includegraphics[width=.3\linewidth]{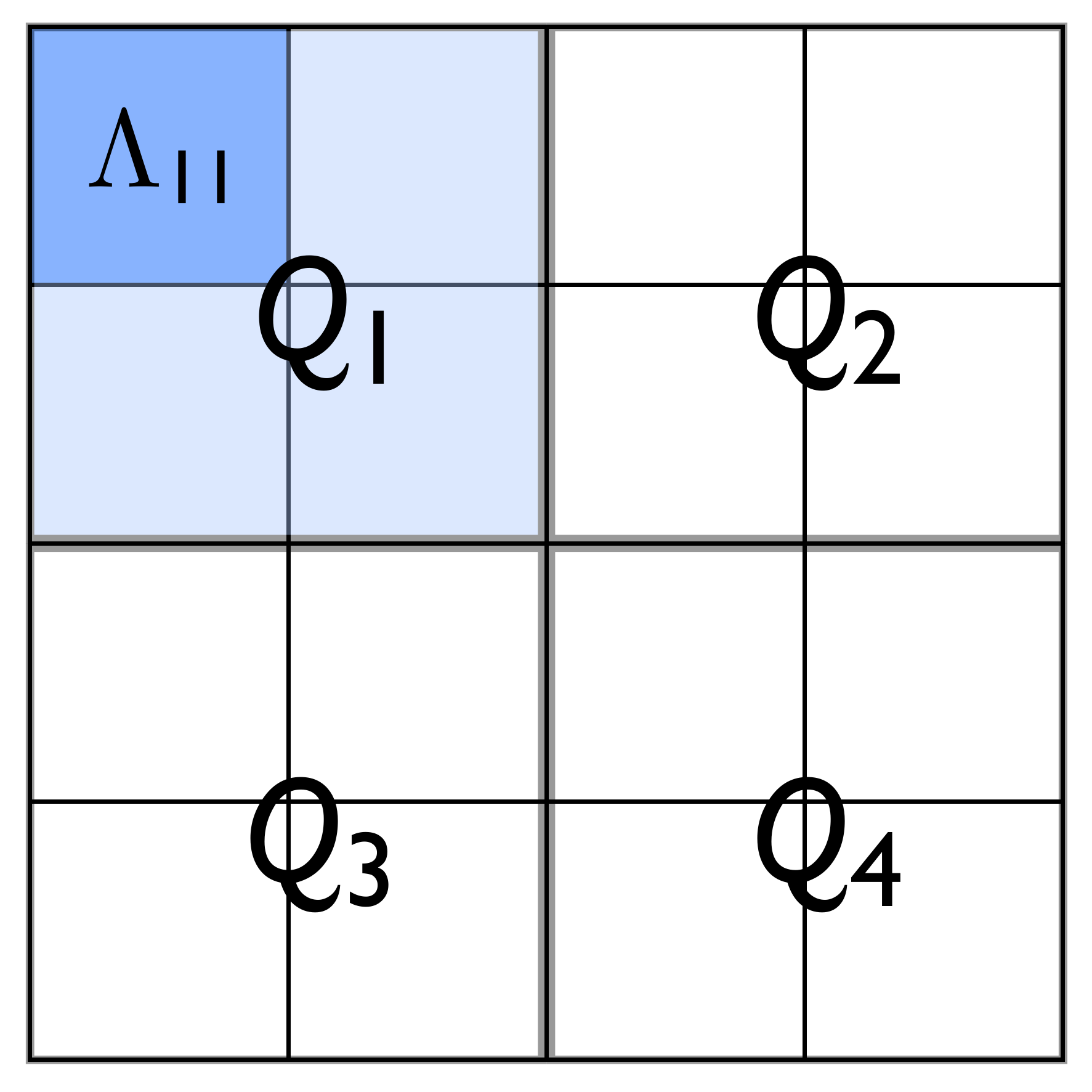}
\caption{A schematic representation of the multiscale decomposition for an image consisting of a $4 \times 4$ grid of pixels. For instance, $\Lambda_{11} = \phi_{11} \phi_{211}$, where $\phi_{11}$ is the proportion expected in the first quadrant of the expected total count across the entire image, i.e., $\phi_{11} = \sum_{j \in Q_1} \Lambda_{1j}$, and $\phi_{211}$ is the proportion of the expected counts in the first quadrant expected in its first pixel, i.e., $\phi_{211}=\Lambda_{11} / \sum_{j \in Q_1} \Lambda_{1j}$.
}
\label{fig:multiscale}
\end{figure}

The relative intensity  of the added component,  $\bLam_1$, is unknown and must be estimated from the data. Following \citet{esch:etal:04} and \citet{conn:vand:07}, we place a multiscale smoothing prior distribution on $\bLam_1$; see also \citet{nowak:kolaczyk:00}. This prior distribution provides a flexible class of models while ensuring stability in the fit.  We illustrate the structure of the prior distribution  by considering an image composed of a simple $4\times 4$ grid of pixels, as in Fig.~\ref{fig:multiscale}. First, we reparameterize $\bLam_1$ using the decomposition
\begin{equation}\label{eq:decomp}
\Lambda_{1i} = 
\l(\sum_{j \in Q_{k(i)}} \Lambda_{1j}\r)
\l(\frac{\Lambda_{1i}}{\sum_{j \in Q_{k(i)}} \Lambda_{1j}}\r),	
\end{equation}
where $k(i)$ indicates which quadrant of the image contains pixel $i$, and $Q_{k(i)}$ is the collection of pixels in that quadrant. This formulation allows us to hierarchically specify the multiscale smoothing prior distribution. At the first level of the hierarchy we model the proportion of the expected total count in each of the four quadrants of the added component. Let $\bphi_1 = (\phi_{11}, \ldots, \phi_{14})$ represent these proportions, i.e., 
\begin{equation}\label{eq:proportions-scale1}
\phi_{1k} = \sum_{j \in Q_{k}} \Lambda_{1j}, \quad k = 1, \ldots, 4.	
\end{equation}
The first subscript of $\phi$ represents the level of the hierarchy, here a one, and the second represents the quadrant number; see Fig.~\ref{fig:multiscale}.

We formulate the prior distribution to encourage fitted values of the expected quadrant proportions $\bphi_1$ that are similar to each other, thus encouraging smoothing in the added component. Mathematically, this is accomplished using a Dirichlet prior distribution,\footnote{In our representation, a (four dimensional) symmetric Dirichlet distribution with parameter $\dirpar$ has probability density function ${\rm pdf}(x_1, x_2, x_3, x_4) = \frac{\Gamma(4\dirpar)}{\Gamma(\dirpar)^4} \prod_{i=1}^4x_i^{\dirpar-1}$; the mean of $x_i$ is $1/4$; the standard deviation of $x_i$ is $\sqrt{3/\{16(4\dirpar+1)\}}$; and the correlation of $x_i$ and $x_j$ is $-1/3$.}
$$
\bphi_1 \sim \text{Dirichlet}\{(\dirpar_1,\dirpar_1,\dirpar_1,\dirpar_1)\}.
$$  
Under this distribution, the larger $\dirpar_1$ is, the smoother the reconstruction is at this level of the hierarchy/resolution in the image; for this reason $\dirpar_1$ is called a {\it smoothing parameter}. Similarly, at the second level of the hierarchy/resolution, we model the expected pixel counts within each quadrant as a proportion of the total expected quadrant count, i.e., we model
\begin{equation}\label{eq:proportions-scale2}
\phi_{2k\ell} =  {\Lambda_{1i} \over \phi_{1k}} =
\frac{\Lambda_{1i}}{\sum_{j \in Q_k} \Lambda_{1j}}, \quad \text{$i =$ $\ell$th pixel in $k$th quadrant};	
\end{equation}
see Fig.~\ref{fig:multiscale}.  Again, we use a Dirichlet distribution: 
$\bphi_{2k} \sim \text{Dirichlet}\{(\dirpar_2,\dirpar_2,\dirpar_2,\dirpar_2)\}$, $k=1,\ldots,4$, where $\bphi_{2k} = (\phi_{2k1}, \ldots, \phi_{2k4})$ with subscripts representing the level of resolution, the quadrant, and the pixel within quadrant. We may use a different smoothing parameter in this level of the hierarchy than in the first level (i.e., $\dirpar_1$ may differ from $\dirpar_2)$ to allow for different structures at different image resolutions. 

For larger images, we can continue this hierarchy using different smoothing parameters for the Dirichlet distribution at different levels of resolution. 
In this way, we might expect little smoothing at the lowest level of resolution and more smoothing at higher levels. A small value of $\dirpar_1$ minimizes smoothing across the four quadrants of the image, while larger values of $\dirpar_k$ for $k>1$ encourage more smoothing at level $k$ of the hierarchy. \citet{esch:etal:04} suggests using cycle spinning to prevent visual artifacts that arise from a fixed multiscale decomposition. 
Cycle spinning consists of randomly translating the origin of the multiscale grid while iteratively updating parameter estimates; for details, see \citet{esch:etal:04}.

\citet{esch:etal:04} apply this  hierarchical prior distribution to derive fitted Bayesian X-ray images in the absence of a baseline model. They 
include a hyperprior distribution to fit the smoothing parameters $\bdirpar = (\dirpar_1, \ldots, \dirpar_D)$, where $D$ is the number of scales in the multiscale decomposition. This strategy alleviates the need to specify the values of the smoothing parameters. We follow their recommendation and use the hyperprior distribution
$\pi(\bdirpar) \propto \prod_{i=1}^D \exp(-1000 \dirpar_i^3),$
which encourages small values of $\dirpar_i$, and so imposes less smoothing, but is not so heavily concentrated near zero as to cause numerical problems. Using this specification of the added component, we confine attention to images that are cropped to $2^D\times 2^D$ pixels for some integer $D$.

Because of their different roles in our model, we place different prior distributions on $\tau_1$ and $\tau_0$. First, $\tau_1$ specifies the total expected count from the added component, and its prior distribution must be flexible enough to allow for values near zero if the baseline model is adequate and for large values if the baseline model is not adequate. We accomplish this using a Gamma distribution\footnote{A Gamma distribution with shape parameter $a$ and rate parameter $b$ has probability density function ${\rm pdf}(x) = \frac{b^a}{\Gamma(a)} x^{a-1} e^{-bx}$, mean $a/b$, and standard deviation $\sqrt{a}/b$.} 
with mean and standard deviation equal to 20. This distribution exhibits significant skewness, with substantial probability near zero and appreciable probability extending to large values.

Second, $\tau_0$ specifies the total expected count from the baseline component. In practice, the observed image typically provides plenty of information to constrain $\tau_0$, since we usually observe at least 100 counts across the entire image and the baseline component is a reasonable description of at least some major image features. Thus, we use a relatively diffuse prior distribution, specifically, the improper distribution, $\pi(\tau_0) \propto \tau_0^{0.001 - 1}$ (it can be shown that under very mild conditions, the posterior distribution will be proper when $\pi(\tau_0) \propto \tau^{\epsilon-1}$ for any $\epsilon > 0$).

\subsection{The null and alternative hypotheses}
\label{sec:hyp}
We are interested in comparing two models for the image. The first corresponds to the hypothesis that the baseline component fully represents the image and no added structure is needed.  The second hypothesis stipulates that the baseline component is insufficient and there is significant structure in the image that can be represented by the added component.  We refer to these two hypotheses as the {\it null hypothesis} and the {\it alternative hypothesis}, respectively.  (Up until now we have referred to these two models as the baseline and full models, respectively. From here on we will employ the more formal terminology, i.e., null and alternative hypothesis/model.) Statistically our goal is to quantify the evidence in the image for deciding between these hypotheses. This choice can be formalized using the notation of Section~\ref{sub:full}: the alternative hypothesis is specified in Equation~\ref{eq:detected} and the null hypothesis arises as the special case where  $\tau_1=0$. Thus, our hypotheses are
\begin{align}
  H_0: \tau_1 &= 0 \label{eq:null_hyp} \\
  H_A: \tau_1 &\sim \pi(\tau_1), \label{eq:alt_hyp}
\end{align}
where $\pi(\tau_1)$ is the prior distribution for $\tau_1$ (under the alternative hypothesis).

In the example of Section~\ref{sub:example} and Fig.~\ref{fig:images}(b), under the null hypothesis, the image is assumed to have an underlying intensity that consists of the baseline component of a flat background and a single point source representing a quasar with no jet. An image of the baseline component (not shown) would look like Fig.~\ref{fig:images}(a) without the two fainter point sources. Under the alternative hypothesis, the assumed underlying intensity is a weighted sum of the quasar-only baseline component and the added multiscale component that allows for additional structure beyond the quasar point source.

The likelihood function and the posterior distribution under the alternative hypothesis, i.e., $H_A$ in Equation~\ref{eq:alt_hyp}, are described in Equations~\ref{eq:bayes}--\ref{eq:newbay}. Similarly, we let
$$
\like_0(\by\obs \mid \btheta_0) = \like(\by\obs \mid \tau_1=0, \btheta_0)
$$
and 
$$
\pi_0(\btheta_0 \mid \by\obs) \propto  \like_0(\by\obs \mid \btheta_0) \pi(\btheta_0)
= \like_0(\by\obs \mid \btheta_0) \pi(\tau_0) \pi(\bnu)
$$
denote the likelihood function and posterior distribution, respectively, under the null hypothesis, i.e., $H_0$ in Equation~\ref{eq:null_hyp}. (Note that if $\tau_1 = 0$, then the likelihood function does not depend on $\bLam_1$ and we do not attempt to estimate it.) 

When fitting the alternative model, sometimes we fix the nuisance parameters of the baseline component, $\bnu = \hat{\bnu}$, perhaps estimating them in a preliminary analysis. In this case,
we work with the conditional posterior distribution of  $(\tau_0, \btheta_1)$ given $\hat{\bnu}$ rather than the full posterior distribution $\pi(\btheta \mid \by\obs)$. 
We denote this conditional posterior distribution  $\pi(\tau_0, \btheta_1 \mid \by\obs, \hat{\bnu})$. When there is little posterior uncertainty in $\bnu$, this ``plug-in posterior distribution'' approximates the marginal posterior distribution, $\pi(\tau_0, \btheta_1 \mid \by\obs) = \int \pi(\btheta \mid \by\obs) {\rm d}\bnu$. 
This is the approach taken in \citet{mcke:etal:15}, where the location and amplitude of each quasar as well the intensity of a uniform background are fit using Sherpa \citep{freeman:etal:01} in separate preliminary analyses; these fitted values are then used to set the relative intensities, $\bLam_0(\hat{\bnu})$. Finally, we fit the alternative model in LIRA, conditioning on $\bLam_0(\hat{\bnu})$, but leaving the scale factor $\tau_0$ as a free parameter.

\section{Testing for structure}\label{sec:testing}

\subsection{Statistical hypothesis testing}
\label{sec:framework}

Although we employ Bayesian methods for model fitting, we consider the classical {\it hypothesis testing} paradigm for model selection.  The test is conducted using a test statistic, denote by $T(\by\obs)$, which is chosen so that larger values of $T(\by\obs)$ are indicative of an added component in the image.  In particular, larger values are less likely to have been obtained as a random fluctuation under the null hypothesis. Thus, if $T(\by\obs)$ is large enough we decide  there is sufficient evidence to conclude that the null hypothesis is inappropriate and there is added structure in the image beyond the baseline component. In this framework, we must determine a threshold for $T(\by\obs)$ such that values of $T(\by\obs)$ greater than the threshold are sufficient evidence to declare detection of structure beyond the baseline component. In the hypothesis testing framework, this is done by limiting the probability of a false detection. Thus, the detection threshold, $T^\star$, is the smallest value such that
\begin{equation}
\Pr\big(T(\by_0) \geq T^\star \mid  \tau_1=0, \btheta_0\big) \leq \alpha, 
\label{eq:typeI}
\end{equation}
where $\by_0$ is a random replicate image generated under the null hypothesis and $\alpha$ is the maximum allowed probability of a false detection.  

Conversely, we can compute the probability under the null hypothesis of observing an image as extreme or more extreme than the observed image, as quantified by the test statistic, i.e., 
\begin{equation}
p = \Pr\big(T(\by_0) \geq T(\by\obs)  \mid  \tau_1=0, \btheta_0\big).
\label{eq:pvaluedef}
\end{equation}
This is called a {\it p-value} and small values, e.g., less than 0.05 or 0.01, are taken as evidence that the image was not generated under the null hypothesis and thus are generally interpreted as evidence in favor of structure in the image beyond the baseline component. Although very popular in practice, p-values are criticized on theoretical grounds from both frequentist and Bayesian perspectives \citep[e.g.,][]{berg:dela:87,zhao:14}. 

\subsection{The test statistic}
\label{sec:teststat}

Before we can compute the p-value in Equation~\ref{eq:pvaluedef}, we need to choose a test statistic $T(\by\obs)$. For a test statistic to be useful, it should provide discrimination between the null and alternative hypotheses. To motivate our choice of test statistic, consider the parameter
\begin{equation}\label{eq:fraction_intensity}
 \xi = \tau_1 / (\tau_1 + \tau_0),
\end{equation}
the proportion of the total image intensity that is due to the added component. If the baseline component fits the data poorly, we expect more of the observed counts to be attributed to the added component, corresponding to large values $\xi$.
On the other hand, if the data are generated under the null hypothesis with $\tau_1 = 0$, we expect more of the observed counts to be attributed to the baseline component, corresponding to $\xi$ near zero. (Formally, under the null hypothesis, $\xi=0$. Nonetheless, its fitted value under the alternative hypothesis will typically be small but positive even if the data are generated under the null hypothesis.)
Thus, $\xi$ is a good candidate for discriminating between the null and alternative hypotheses.

Unfortunately, $\xi$ is a parameter, not a statistic; that is, it cannot be computed directly as a function of the data $\by\obs$. However, the posterior distribution of $\xi$ under the alternative hypothesis, conditional on the data, can be computed from the data. This motivates us to use a feature of the posterior distribution of $\xi$ as a test statistic. In particular, our test statistic is a posterior tail probability of $\xi$.
Given a threshold $c$, we let
\begin{equation}\label{eq:test_stat}
  T_{c}(\by\obs) = \Pr(\xi \geq c \mid \by\obs),
\end{equation}
where the probability is taken with respect to $\pi(\btheta \mid \by\obs)$, the posterior distribution under the alternative hypothesis. 
To some readers, it may seem more natural to use the fitted value of $\xi$ as a test statistic, but as we discuss in Section~\ref{sub:comp}, there are 
advantages to using the tail probability, $T_c(\by\obs)$. This choice allows us to treat $c$ as a tuning parameter and thereby to select a more powerful test statistic than the fitted value of $\xi$. 
Although $T_{c}(\by\obs)$ involves computations under the posterior distribution, it is a true statistic in that (for a fixed prior distribution and a given $c$) it is a function only of the data, and not of any unknown parameters.
Computing $T_{c}(\by\obs)$ with respect to the plug-in posterior distribution $\pi(\btheta \mid \by\obs, \hat{\bnu})$ also leads to a valid test statistic in the same sense.\footnote{ 
Indeed, the resulting test statistic is valid whether or not $\pi(\btheta\mid\by\obs, \hat{\bnu})$ is a good approximation of $\pi(\btheta \mid \by\obs)$, because the posterior distribution $\pi(\btheta\mid\by\obs, \hat{\bnu})$ can in principle be computed as a function of the data. 
See Section \ref{sec:mcmcissues} for discussion of the trade-offs involved when using computational approximations of test statistics.
}

When the alternative hypothesis is true, we expect large values of $\xi$ to explain the data better than small values, and we therefore expect a high posterior probability that $\xi$ exceeds an appropriate value of $c$; that is, we expect a large value of $T_c(\by\obs)$. Conversely, when the null hypothesis is true, $T_c(\by\obs)$ is typically small because small values of $\xi$ tend to better explain the data. Thus, $T_c(\by\obs)$ behaves differently under the null and alternative hypotheses, making it a reasonable candidate for a test statistic to discriminate between the two hypotheses.

It can be useful to substitute other choices of $\xi = h(\btheta)$ in Equation~\ref{eq:test_stat} to define different test statistics $T_c(\by\obs)$. 
For instance, often we are interested detecting departures from the baseline model in a specific, known region of the image; see \citet{mcke:etal:15} for numerous examples. In this case, it is possible to design a test statistic targeted at the given region. Let $R$ be a collection of pixel indices defining a region of interest on the image. To test whether there is a significant departure from the null hypothesis in the region $R$, we can use as a test statistic a posterior tail probability of the parameter
\begin{equation}\label{eq:fraction_intensity_local}
 \xi_R = \frac{\sum_{j \in R} \tau_1 \Lambda_{1j}}{\sum_{j \in R} (\tau_1  \Lambda_{1j}  + \tau_0 \Lambda_{0j})},
\end{equation}
the fraction of the total intensity in $R$ attributed to the added component. In particular, our test statistic is 
\begin{equation}\label{eq:test_stat_local}
	T_{R,c}(\by\obs) = \Pr(\xi_R \geq c \mid \by\obs).
\end{equation}
Because it only considers pixels in $R$, $T_{R,c}(\by\obs)$ only has power to detect departures from the null hypothesis that manifest in the region of interest. However, by ignoring regions of the image where little or no departure from the null hypothesis is expected, $T_{R,c}(\by\obs)$ may be more powerful than the image-wide $T_c(\by\obs)$ for detecting departures concentrated in $R$.

\subsection{A fully specified null hypothesis}
\label{sec:nonuissance}

Because the probabilities in Equations~\ref{eq:typeI}--\ref{eq:pvaluedef} depend on the unknown parameters $\btheta_0$, the probability of a false detection and hence $T^\star$  and the p-value cannot be computed. This is a nuisance and why parameters that are unknown under the null hypothesis, like $\tau_0$ and $\bnu$, are called {\it nuisance parameters}. We set aside this difficulty for the moment to focus on statistical issues that arise in the absence of nuisance parameters, but return to it in Section~\ref{sub:nuisance}.

In particular, we start by assuming that $\btheta_0$ is known, the null hypothesis has no unknown parameters, and $\like_0(\by\obs \mid \btheta_0) = \like_0(\by\obs)$. 
In practice, $\btheta_0$ is never known exactly, but in some situations we may be able to estimate it with high enough precision that we can treat it as fixed and known. However, great care must be taken when fixing $\btheta_0$.  If it is fixed at an inappropriate value, evidence against the null hypothesis may not indicate that the null model is inappropriate so much as that the fixed values of $\btheta_0$ are inappropriate. For example, $\tau_0$  should not be fixed at an  arbitrary value.  If $\tau_0 \bLam_0(\bnu)$ is fixed, then the expected total count under the null hypothesis is also fixed. If this expected total count differs significantly from the observed total count, the null hypothesis can be rejected on this basis alone.
Generally speaking, if $\tau_0$ is to be fixed, it should be set equal to a reasonable estimate of the total expected count after adjusting for detector inefficiencies, for example to  
\begin{equation}
\hat\tau_0 =  \sum_{i=1}^n y_i  \Big/ \sum_{i=1}^n \sum_{j=1}^n P_{ij}A_j \Lambda_{0j}(\hat\bnu),
\label{eq:esttau0}
\end{equation}
and this should only be done if the total count is large. 
For example, when testing for a quasar jet, we fit the location and amplitude of the point source and the intensity of a constant background using Sherpa \citep{freeman:etal:01}. In Section~\ref{sec:numerical} and \citet{mcke:etal:15}, we use the fitted values of these parameters to fix $\tau_0 \bLam_0(\bnu) = \hat{\tau}_0 \bLam_0(\hat{\bnu})$ in the null model.  When fitting the alternative model, however, we recommend never fixing $\tau_0$, because this would leave no flexibility to reduce the emission attributed to the baseline component and increase the emission attributed to the added multiscale component. Instead, as described at the end of Section~\ref{sec:hyp}, under the alternative model, we fix $\bnu = \hat{\bnu}$ according to the fitted values from Sherpa, but allow $\tau_0$ to be estimated.

With $\xi$ defined in Equation \ref{eq:fraction_intensity}, larger values of $T_c(\by)$ are more unusual under the null hypothesis. The p-value in Equation~\ref{eq:pvaluedef} simplifies to 
\begin{equation}\label{eq:pvalue}
  p = {\Pr}\l(T_c(\by_0) \geq T_c(\by\obs) \mid \tau_1=0\r),
\end{equation}
where $\by_0$ is a randomly generated image under the null hypothesis and $\by\obs$ is the fixed observed image. Because there are no nuisance parameters, this p-value can in principle be computed exactly.

\subsection{Computing the statistical significance}\label{sub:comp}

The primary advantage of using a tail probability of $\pi(\btheta \mid \by\obs)$ rather than a fitted value of $\xi$ or the likelihood ratio test as the test statistic is computational.  We describe this advantage here.

Although we cannot directly evaluate the tail probability $T_c(\by\obs) = \Pr(\xi \geq c \mid \by\obs)$ under the alternative model, we can estimate it numerically via MCMC. Even if we could compute $T_c(\by\obs)$ directly, we would need to compute the probability in Equation~\ref{eq:pvaluedef} to evaluate the p-value, and this too is most easily obtained through 
Monte Carlo simulation.

MCMC involves obtaining  $L$ correlated draws $\xi^{(1)}\obs,\ldots, \xi^{(L)}\obs$ from the posterior distribution under the alternative model, $\pi(\xi \mid \by\obs)$. This can be accomplished using the LIRA package, which relies on the Gibbs sampling algorithm described in \citet{esch:etal:04}. Specifically, LIRA delivers a correlated sample $\btheta^{(1)}\obs,\ldots,\btheta^{(L)}\obs$ from the full posterior distribution $\pi(\btheta \mid \by\obs)$, and we then compute each $\xi^{(\ell)}\obs$ as a function of each $\btheta^{(\ell)}\obs$, say $\xi^{(\ell)}\obs = h(\btheta^{(\ell)}\obs)$. With the MCMC sample in hand we  estimate $T_c(\by\obs)$ as
\begin{equation}
\label{eq:mcT}
\widehat{T}_c(\by\obs) = \frac{1}{L} \sum_{\ell=1}^L 1\{\xi^{(\ell)}\obs \geq c\},
\end{equation}
where $1\{\cdot\}$ is the indicator function that equals one if its argument is true and is zero otherwise. 

A straightforward way to estimate the p-value is to simulate $M$ independent replicate images under the null hypothesis and then fit the alternative model and compute the test statistic for each. This can be accomplished via the following method.

\medskip
\noindent
{\bf Direct P-value Method:}

\noindent
For $j=1,\ldots,M$,
\begin{enumerate}
\item Simulate  $\by_0^{(j)} \sim \like_0(\by_0)$;
\item Fit the alternative model to  $\by_0^{(j)}$ by running LIRA to obtain $L$ correlated draws $\xi^{(j,1)},\ldots, \xi^{(j,L)}$ from $\pi(\xi \mid \by_0^{(j)})$; and 
\item Compute the estimated test statistic $\widehat{T}_c(\by_0^{(j)})$ using Equation~\ref{eq:mcT} with $\xi^{(\ell)}\obs$ replaced with $\xi^{(j,\ell)}$.
\end{enumerate}

 \noindent
Finally, estimate the p-value with the Monte Carlo p-value,
\begin{equation}\label{eq:mcpvalue}
    \hat{p} = \frac{1 + \sum_{j=1}^M 1\l\{\widehat{T}_c(\by_0^{(j)}) \geq \widehat{T}_c(\by\obs)\r\}}{1 + M},
\end{equation}
recommended in \citet{davison:hinkley:97}. Equation \ref{eq:mcpvalue} adds one to the numerator and denominator of the naive Monte Carlo p-value,
\begin{equation}\label{eq:naivemcpvalue}
\hat{p}_{{\rm naive}} = \frac{\sum_{j=1}^M 1\l\{\widehat{T}_{c}(\by_0^{(j)}) \geq \widehat{T}_{c}(\by\obs)\r\}}{M}.
\end{equation}
We use $\hat{p}$ instead of $\hat{p}_{{\rm naive}}$ to guarantee that a testing procedure that rejects the null hypothesis when the p-value is less than or equal to a pre-specified $\alpha$ has false positive rate no greater than $\alpha$. The naive Monte Carlo p-value $\hat{p}_{{\rm naive}}$ does not control the false positive rate in this manner.\footnote{Under the null hypothesis, $\Pr(\hat{p}_{{\rm naive}} = i/M) = 1/(M+1)$, for $i=0,1,\ldots,M$. Thus, the true false positive rate using the naive Monte Carlo p-value is $\Pr(\hat{p}_{{\rm naive}} \leq \alpha) = (\lfloor M\alpha \rfloor +1)/(M+1)$, which is greater than $\alpha$ for some choices of $\alpha$ (such as $\alpha = j/M$, if $j \in \{0, 1, \ldots, M-1\}$).}
For further discussion and a numerical demonstration, see Section~\ref{sub:power}. 

Fitting our Bayesian imaging model via MCMC using LIRA is computationally expensive. Unfortunately, the Direct P-value Method requires us to run LIRA $M$ times and $M$ must be very large to have any chance of achieving a high level of statistical significance (i.e., a low value of $\hat{p}$) because $\hat{p} \geq 1/(1+M)$. This requires devoting $M$ times the computational resources to analyzing simulated replicate images as analyzing the observed image. This is very computationally expensive and often unacceptable in applied work. In practice, often the reconstructed image under the alternative model is of primary interest and the statistical p-value is intended  as an additional check to prevent over-interpreting apparent structure in the noise as discovery of real structure in the signal. Although the p-value is often of secondary interest, preventing over-interpretation of images is of course critically important.  Nonetheless, devoting 1000 times (or more) the computing time to computing a p-value is often infeasible.

\begin{figure}[t]
\centering
\includegraphics[width=.8\linewidth]{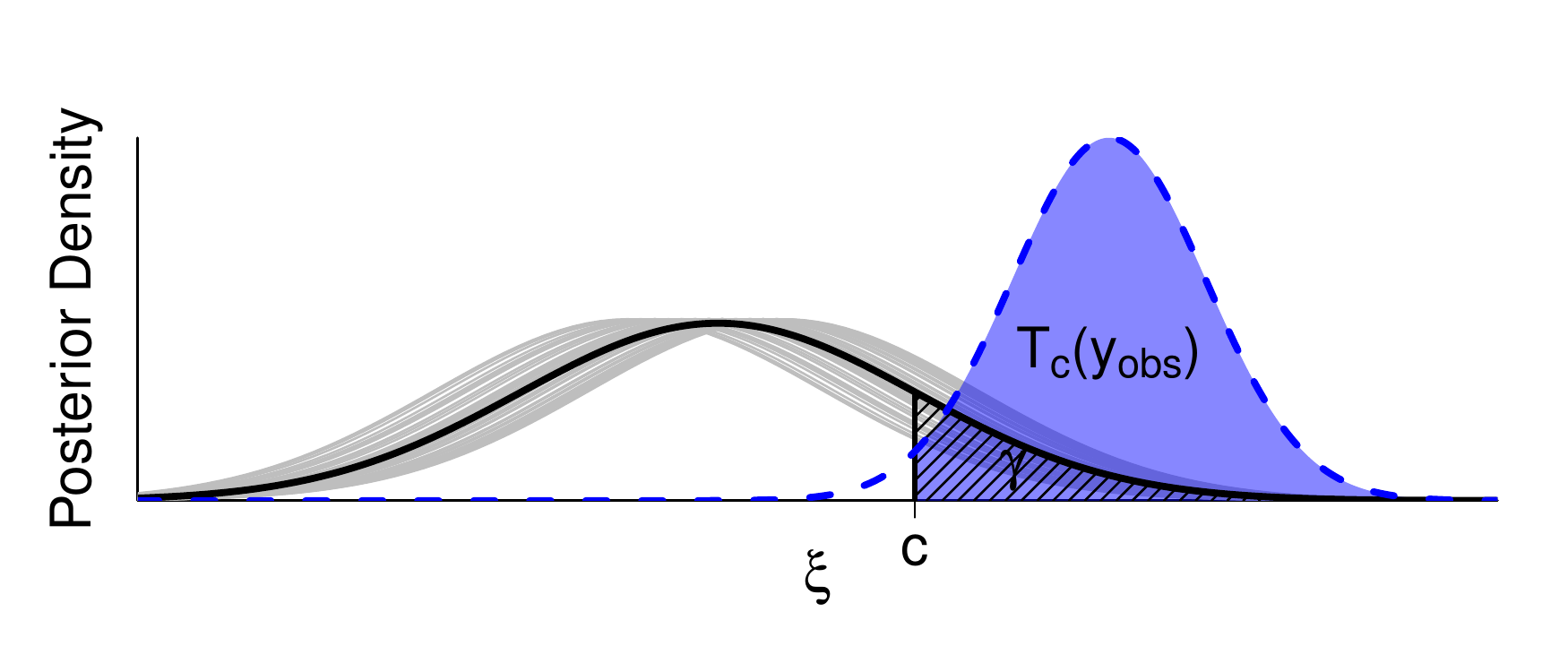}
\caption{
	A schematic illustration of the quantities used to compute the upper bound $u$ in Equation~\ref{eq:Markov}. The gray solid lines represent posterior densities $\pi(\xi \mid \by_0^{(j)})$ given a sample of images $\by_0^{(j)}$ simulated under the null; the black solid line is $g(\xi)$, computed as the average of the posterior densities shown in gray; and the blue dashed line is the posterior density $\pi(\xi \mid \by\obs)$ given an observed image $\by\obs$. To compute $\hat{u}$, we fix $\gamma$ (the hatched area under $g(\xi)$ to the right of $c$); compute $c$ as the $(1-\gamma)$ quantile of $g(\xi)$, i.e., $\int_c^1 g(\xi) {\rm d}\xi = \gamma$, from an MCMC sample from $g(\xi)$; and use this value of $c$ to compute $T_c(\by\obs)$ (the area under $\pi(\xi \mid \by\obs)$ to the right of $c$, shaded in blue) from an MCMC sample from $\pi(\xi \mid \by\obs)$.
}
\label{fig:tailprob}
\end{figure}

To reduce the computational requirements of this significance test, we propose estimating not the p-value but an upper bound on it. In Appendix~\ref{sec:technical} we show that
\begin{equation}\label{eq:Markov}
      p \leq \frac{\gamma}{T_c(\by\obs)} = u,
\end{equation}
where $\gamma = \Pr(\xi \geq c)$ under the distribution 
\begin{equation}\label{eq:expost}
g(\xi) =
E\l\{\pi(\xi \mid \by_0) \r\} \\
= \sum_{\by_0} \pi(\xi \mid \by_0) \, \like_0(\by_0),   
\end{equation}
the expectation under the null hypothesis of the posterior distribution under the alternative hypothesis. To compute the upper bound, $u$, on the p-value for a given fixed value of $\gamma$, the denominator of Equation~\ref{eq:Markov} must be estimated, and doing so involves two sources of uncertainty: (i)  estimating the quantile $c$  of $g(\xi)$ and (ii) estimating the tail probability under the alternative hypothesis given $\by\obs$; see Fig.~\ref{fig:tailprob}. The upper bound can be computed using the following method.

\medskip
\noindent
{\bf Upper Bound Method:}

\noindent
For $j=1,\ldots,M$,
\begin{enumerate}
\item Simulate  $\by_0^{(j)} \sim \like_0(\by)$;
\item Fit the alternative model to  $\by_0^{(j)}$ by running LIRA to obtain $L$ correlated draws $\xi^{(j,1)},\ldots, \xi^{(j,L)}$ from $\pi(\xi \mid \by_0^{(j)})$;
\end{enumerate}

\noindent
Then, set $\hat{c}$ equal to the estimated $(1-\gamma)$ quantile of $g(\xi)$, using the $LM$ posterior draws $\xi^{(j,\ell)}$, $j=1,\ldots,M$; $\ell = 1, \ldots, L$. For instance, $\hat{c}$ may be set equal to the $(LM\gamma)$th largest  value among all of the $\xi^{(j,\ell)}$. Finally, compute the estimated test statistic $\widehat{T}_{\hat{c}}(\by\obs)$ using Equation~\ref{eq:mcT} with $c$ replaced with the estimate $\hat{c}$. Our estimate of the upper bound is
\begin{equation}\label{eq:u_hat}
\hat{u} = \gamma / \widehat{T}_{\hat{c}}(\by\obs).
\end{equation}

Equation~\ref{eq:u_hat} reveals one of the advantages of using a posterior tail probability as a test statistic. If we were to replace Equation~\ref{eq:fraction_intensity} with another choice of (an always non-negative) $\xi$,  we could use its fitted value as a test statistic and derive an upper bound on the appropriate p-value as in Appendix~\ref{sec:technical}, but this upper bound could be quite large and there would be no remedy. However, using the tail probability $T_c(\by)$ as a test statistic allows us flexibility in the choice of the tuning parameter $c$. We use this flexibility to fix the numerator in Equation~\ref{eq:u_hat} to a reasonable small value (for more on the choice of $\gamma$, see Section~\ref{sub:practical}), enabling the possibility of obtaining a small upper bound. Whereas $\hat{p} \geq 1/(1+M)$, it is possible to obtain a $\hat{u}$ much less than $1/(1+M)$ if $\gamma$ is chosen appropriately. 
{\it This is the primary advantage of the Upper Bound Method. It allows us to establish statistical significance with fewer simulated replicate images and thus can be appreciably faster in practice. 
}

\subsection{Null hypothesis with unknown parameters}\label{sub:nuisance}

In this section, we consider settings in which the parameters of the null model have non-negligible uncertainties. When such uncertainty exists, fixing the null model by substituting estimates for these unknown parameters can lead to problems. As mentioned in Section~\ref{sec:nonuissance}, we might reject the null hypothesis because the parameters have been fixed at inappropriate values, not because the null model itself is incorrect.

When there is non-negligible uncertainty in the parameters of the null model, $\like_0(\by_0 \mid \btheta_0)$ depends on $\btheta_0$. We assume that under the null hypothesis, we can obtain the posterior distribution $\pi_0(\btheta_0 \mid \by\obs)$ of these nuisance parameters under a prior $\pi_0(\btheta_0)$. With this approach, the p-value in Equation \ref{eq:pvalue} can be calculated for each fixed value of $\btheta_0$. We denote this
\begin{equation}
  p(\btheta_0) = {\Pr}\l(T_c(\by_0) \geq T_c(\by\obs) \mid \btheta_0 \r),
\end{equation}
where the probability is taken over the sampling distribution $\like_0(\by_0 \mid \btheta_0)$ for a fixed value of the parameter $\btheta_0$. A Bayesian posterior predictive p-value 
(ppp-value) averages $p(\btheta_0)$ over the posterior for $\btheta_0$ under the null hypothesis and is given by
\begin{align}\label{eq:ppp}
  \textrm{ppp-value} &= E\l\{ p(\btheta_0) \mid \by\obs\r\} \\
  &= \int p(\btheta_0) \ \pi_0(\btheta_0 \mid \by\obs) \ {\rm d}\btheta_0 \nonumber.
\end{align}
See \citet{rubin:84}, \citet{meng:94}, and \citet{gelman:meng:stern:96}, among others, for discussions of the properties of ppp-values.\footnote{
\citet{meng:94} investigates the frequency properties of ppp-values under the prior predictive distribution $\int \like_0(\by_0 \mid \btheta_0) \ \pi_0(\btheta_0) \ {\rm d}\btheta_0$. Under this  distribution, the ppp-value is more concentrated around 0.5 than a uniform distribution, and a test that rejects the null hypothesis when the $\textrm{ppp-value} \leq \alpha$ will have false positive rate no greater than $2 \alpha$. Similarly, \citet{sinharay:stern:03} and \citet{bayarri:castellanos:07} discuss the conservativeness of ppp-values in hierarchical models.}

We can compute an upper bound on the ppp-value using an expression similar to Equation~\ref{eq:Markov} 
\begin{equation}\label{eq:Markovppp}
	\textrm{ppp-value} \leq \frac{\gamma}{T_c(\by\obs)} = u_{\rm ppp},
\end{equation}
where now $\gamma = \Pr(\xi \geq c \mid \by\obs)$ under the distribution\footnote{The distribution $\int \like_0(\by_0 \mid \btheta_0) \ \pi_0(\btheta_0 \mid \by\obs) \ {\rm d}\btheta_0$ is the {\it posterior predictive} distribution of $\by_0$ given $\by\obs$ under the null hypothesis, so $g(\xi \mid \by\obs)$ is the posterior predictive expectation under the null hypothesis of the posterior distribution of $\xi$ under the alternative hypothesis.}
\begin{align}
    g(\xi \mid \by\obs) 
    &= E\l\{\pi(\xi \mid \by_0) \mid \by\obs\r\} \nonumber\\
    &= \sum_{\by_0} \pi(\xi \mid \by_0) \int \like_0(\by_0 \mid \btheta_0) \ \pi_0(\btheta_0 \mid \by\obs) \ {\rm d}\btheta_0. \label{eq:expostpred}
\end{align}

To implement the Direct P-value and Upper Bound Methods in this setting, we only need to modify each method's Step 1, replacing it with 
\begin{itemize}
	\item[1$'$.] Simulate $\btheta_0^{(j)} \sim \pi_0(\btheta_0 \mid \by\obs)$, followed by $\by_0^{(j)} \sim \like_0(\by_0 \mid \btheta_0^{(j)})$.
\end{itemize}
This requires, of course, that we can obtain draws from $\pi_0(\btheta_0 \mid \by\obs)$, the posterior distribution of the unknown parameters under the null hypothesis. In practice, we can approximate this posterior distribution using the estimated uncertainties from fitting the baseline component in Sherpa \citep{freeman:etal:01}. The implementation of Step 2, while notationally the same, is more complicated when there are unknown parameters under the null hypotheses because these
parameters must be fit.

\section{Numerical results}\label{sec:numerical}

In this section, we extend the example in Section~\ref{sub:example} and investigate the performance of our proposed method on images, both simulated and real, of quasars with possible jets.

\subsection{Three simulated images}\label{sub:three-images}

We begin by analyzing the simulated image of Section~\ref{sub:example} and compare the results to analyses of two other simulated images, one with a weaker jet and one with a stronger jet. Recall that in the image of Section~\ref{sub:example} the jet was composed of two additional Gaussian components, each with 20 expected counts.  We call this the ``medium jet.'' The simulated images for the weak and strong jets were constructed identically to the medium jet image, except that each additional Gaussian component had 10 expected counts in the weak jet and 35 in the strong jet.

Each ground truth image was convolved with the same PSF to obtain the Poisson intensity in each pixel, with which we sampled to generate the simulated observations. The PSF (Fig.~\ref{fig:images}(c)) was generated with SAOTrace\footnote{{\tt http://cxcoptics.cfa.harvard.edu/SAOTrace/Index.html}} to reflect the observation conditions of the quasar analyzed in Section~\ref{sub:data}, which was observed by the {\it Chandra} X-ray Observatory.

To analyze each simulated image, we created a baseline component $\tau_0\bLam_0$ for each jet strength that consisted of the simulated quasar with the correct location and intensity (analogous to a real data analysis in which these parameters are well constrained by the data and can be treated as known) and a uniform background. The expected count due to background tabulated in $\tau_0\bLam_0$ was set to the sum of the actual background and the jet components, so that the total expected count in the null model was equal to the total in the true image for each jet strength.
This prevents rejection of the null hypothesis purely on the basis of the total count.  
In this simulation the null hypothesis was fully specified with no unknown parameters.
The posterior means of $\tau_1 \Lambda_{1j}$ for the medium jet are shown in Figure~\ref{fig:images}(d), in which the two sources not included in the baseline component are clearly visible.

\begin{figure}[t]
    \centering
    \includegraphics[width=.6\linewidth]{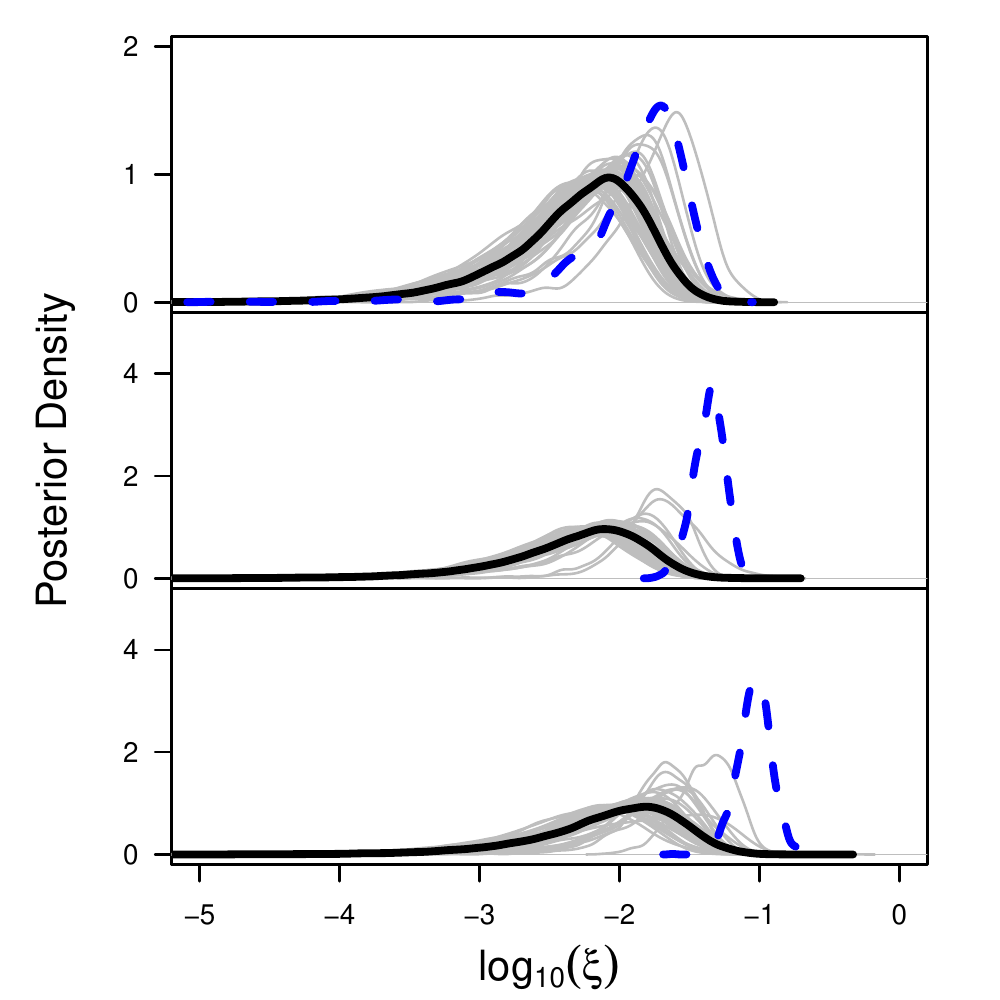}
    \caption{Estimated posterior distributions 
    for the simulated quasar with a weak jet (top panel), simulated quasar with a medium-strength jet (middle panel), and observed quasar discussed in Section~\ref{sub:data} (bottom panel). The dashed blue lines are $\pi(\xi \mid \by\obs)$, the posterior under the alternative hypothesis given the observed data; the light gray solid lines are $\pi(\xi \mid \by_0^{(j)})$, the posterior under the alternative hypothesis given images $\by_0^{(j)}$ simulated under the null hypothesis; and the heavy black solid lines are $g(\xi)$, the expectation under the null hypothesis of the posterior distribution under the alternative hypothesis.}
    \label{fig:posteriors}
\end{figure}

We also simulated $M=50$ images under the null hypothesis for each jet strength. For each simulated and replicate image, we fit the alternative hypothesis via MCMC using the LIRA package, obtaining 1800 draws from the posterior distribution after discarding the initial 200 steps as burn-in. The posterior distributions of $\xi$ for the medium jet are shown in the middle panel of Fig.~\ref{fig:posteriors}. As expected, $\pi(\xi \mid \by\obs)$ tends to be to the right of the $\pi(\xi \mid \by_0^{(j)})$, because a higher fraction of observed counts are attributed to the added component $\tau_1 \bLam_1$ with $\by\obs$ than with most of the  $\by_0^{(j)}$.

\begin{figure}[t]
\centering
\includegraphics[width=.6\linewidth]{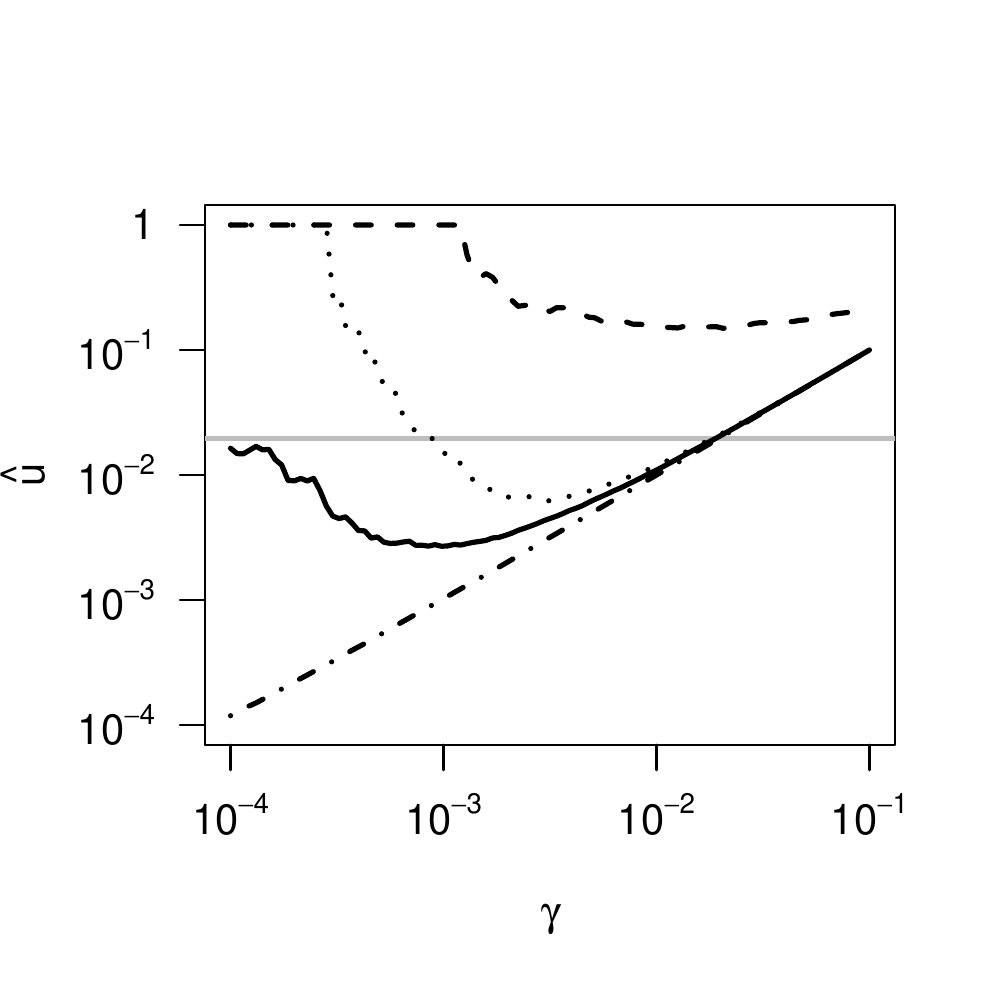}
\caption{Estimated upper bounds $\hat{u}$ for varying values of the upper tail probability, $\gamma$, of $g(\xi)$, for the simulated jets of Section~\ref{sub:three-images} (weak: dashed, medium: dotted, strong: dashed-dotted) and the observed quasar (solid black line) of Section~\ref{sub:data}. The solid gray line is at $1/51$, the minimum achievable $\hat{p}$ as defined in Equation \ref{eq:mcpvalue} when $M=50$.}
\label{fig:upper_bounds}
\end{figure}

Fig.~\ref{fig:upper_bounds} displays the estimated p-value upper bounds, $\hat{u}$, for the simulated images with weak, medium, and strong jets, as a function of the upper tail probability, $\gamma$, of $g(\xi)$. For the weak jet, $\hat{u}$ is consistent with the null hypothesis. However, for the medium and strong jets, if $\gamma$ is in an appropriate interval, $\hat{u}$ appears to reveal significant evidence of inadequacy of the null hypothesis. We discuss bootstrap estimation of the uncertainty in $\hat{u}$ in Section~\ref{sub:data}.

Using the Direct P-value Method, the best (i.e., lowest) p-value that can be achieved with $M=50$ is $\hat{p} = 1/51$. If we were to use $\hat{p}$ as defined in Equation \ref{eq:mcpvalue}, this minimum value would be achieved for the medium simulated jet when $c< 0.06$ and for the strong jet when $c<0.14$. For the weak jet, the smallest $\hat{p}$ that would be achieved is $2/51$, when $c < .04$. The Upper Bound Method can lead to higher significance than the Direct P-value Method for the medium and strong jets, but not for the weak jet. The top panel of Fig.~\ref{fig:posteriors} helps illustrate this, as $\pi(\xi \mid \by\obs)$ for the weak jet is centered slightly to the right of most of the $\pi(\xi \mid\by_0^{(j)})$, but the right tail of $\pi(\xi \mid\by\obs)$ does not extend much past the right tail of $g(\xi)$, preventing the upper bound from being very low.

\subsection{Practical implementation issues}\label{sub:practical}

\subsubsection{Choosing $\gamma$}
As seen in Figure~\ref{fig:upper_bounds}, the Upper Bound Method generates a family of upper bounds $\hat{u}(\gamma)$ corresponding to different choices of $\gamma \in [0,1]$. It would not be valid to simply report the minimum of these upper bounds, $\min_{0 \leq \gamma \leq 1} \hat{u}(\gamma)$, as the statistical significance. This would be analogous to computing multiple p-values and only reporting the most significant (i.e., smallest) one. If multiple tests are performed, each with a bounded probability of returning a false detection, the probability of {\it at least one} false detection among the multiple tests increases with the number of tests.  This phenomenon is often referred to as the {\it look elsewhere effect}  \citep[e.g.,][]{gros:vite:10,vand:14}. Likewise, 
rejecting the null hypothesis if the minimum of multiple p-values is below some threshold $\alpha$ does not guarantee that the probability of false detection is less than $\alpha$. %This is 

Thus, to implement the Upper Bound Method in practice, we need a procedure to choose $\gamma$ in order to eliminate these multiple comparisons problems and allow us to report a single upper bound. Suppose we could compute $T_c(\by)$ exactly, given $c$ and $\by$. This seems a reasonable simplifying assumption if we run MCMC until the effective sample sizes of each posterior sample are sufficiently large and do not use an extreme value of $c$. Under this assumption, the Monte Carlo error in $\hat{u}$ comes exclusively from the uncertainty in $\hat{c}$ due to only having $M$ draws $\by_0^{(1)},\ldots,\by_0^{(M)}$ from $\like_0(\by_0)$. By Equation \ref{eq:u_hat}, $\hat{u} \geq \gamma$, so we should  choose $\gamma$ as small as possible in order to increase the chance of achieving a small $\hat{u}$ and hence a high statistical significance. However, there is a trade-off: the Monte Carlo error in estimating the quantile $c$ grows as $\gamma$ becomes smaller, leading to larger Monte Carlo error in estimating $u$. One  approach to selecting $\gamma$  is to estimate the Monte Carlo error in $\hat{u}$ for a range of values of $\gamma$ and choose the smallest $\gamma$ for which this error value is acceptably small. Choosing $\gamma$ based on the estimated {\it uncertainty} in $\hat{u}$ alleviates some of the multiple comparisons concerns that would arise if we chose $\gamma$ based on the estimated upper bounds themselves. The Monte Carlo error can be estimated via bootstrap by resampling with replacement from $\{\by_0^{(1)}, \ldots, \by_0^{(M)}\}$. We demonstrate this in Section~\ref{sub:data}; see Fig.~\ref{fig:bootstrap}. In our analyses, we found that values of $\gamma$ in the range of 0.005 to 0.01 appeared reasonable when $M=50$.

\subsubsection{Implementation of a suite of MCMC samplers}
\label{sec:mcmcissues}
The reason that we aim to reduce the number, $M$, of replicate images that must be analyzed is not just to save computer time, but also because each replicate must be fit using MCMC, which can be temperamental in practice.  Indeed we can never precisely compute $T_c(\by)$, but only an estimate of the tail probability in Equation~\ref{eq:test_stat} with $\widehat{T}_c(\by)$. Monte Carlo error affects the variance of this estimate and lack of MCMC convergence causes bias. Even a biased estimator, however, can be a valid test statistic, although perhaps a less powerful one.  (A similar view of computational errors when using Monte Carlo to evaluate the sampling distribution of a likelihood ratio test was espoused by Protassov et al. (\citeyear{prot:etal:02}).) The key is that to guarantee a valid statistical test, precisely the same procedure must be implemented for each replicate image as for the observed images.
Thus, the same methods for choosing starting values and diagnosing MCMC convergence must be performed for each image. The danger lies in the temptation to carefully implement and monitor MCMC for the observed image, but not for the $M$ replicate images. Such differences in implementation may result in a systematic bias that may be mistaken as a statistically significant difference, simply because the test statistic function $\widehat{T}_c(\cdot)$ applied to replicate images $\by_0^{(j)}$ is different than the function applied to the observed image $\by\obs$.

\subsection{False positive rate and statistical power}\label{sub:power}

To investigate the statistical properties of our proposed method, we applied our procedure to a total of 6000 simulated observations. Specifically, we simulated $1000$ observed images under each of three null hypotheses (i.e., with no jet, but with differing background rates) and $1000$ observed images under each of the three specific alternative hypotheses  used in the simulations of Section~\ref{sub:three-images} (i.e., with a weak, medium, or strong jet).  The null hypotheses were not identical for the different jet strengths in that the background intensities under the null hypotheses were higher when the alternative contained a stronger simulated jet. All simulated observed images were generated as in Section~\ref{sub:three-images}.  We considered a testing procedure that first chooses a tail probability $\gamma$ of $g(\xi)$ and a nominal significance level $\alpha$, and rejects the null hypothesis if the estimated upper bound $\hat{u} \leq \alpha$, where $\hat{u}$ is defined in Equation \ref{eq:u_hat}. We evaluated this procedure for a range of values of $\gamma$. 

To approximate $g(\xi)$, we sampled  $M=50$ replicate images under each null hypothesis. A full simulation would require fitting the null model to each of the 6000 simulated observations, generating 50 replicate images from the fitted null model of each, and performing a full Bayesian analysis of all 300,000 resulting images.
To reduce the computational demands of this simulation study, we performed an approximate simulation. We generated 500 additional replicate images from each of the three fixed null hypotheses (one for each background rate) and analyzed each replicate image using MCMC. For each simulated observed image, we resampled 50 replicate images without replacement from this collection of 500 replicate images, and used the corresponding 50 MCMC runs to construct $g(\xi)$.

We compared this approach to two Direct P-value Methods. The first uses the \citet{davison:hinkley:97} Monte Carlo p-value of Equation~\ref{eq:mcpvalue} and the second uses the naive Monte Carlo p-value of Equation~\ref{eq:naivemcpvalue}. Both Direct P-value Methods use the same test statistic as the Upper Bound Method and thus require an estimate of $c$. To ensure fair comparisons, for both direct methods and for every value of $\gamma$,  we set $c= \hat{c}$, the estimated $(1-\gamma)$ quantile of $g(\xi)$, and reject the null hypothesis if the estimated p-value is less than or equal to $\alpha$.

Table~\ref{tab:sim-results} presents the estimated false positive rate and statistical power\footnote{
The estimated {\it false positive rate} is the fraction of images simulated under the null hypothesis for which the null hypothesis was incorrectly rejected and thus structure was falsely detected. The estimated {\it statistical power} is the fraction of images simulated under the alternative hypothesis for which the null hypothesis was rejected and thus true structure was detected.
} 
for the three procedures for a range of choices of $\gamma$ and $\alpha$.  Because $\hat{u} \geq \gamma$, it is impossible to reject the null hypothesis using the upper bound approach if $\gamma > \alpha$, so we do not include such choices in our comparisons. 

If we knew the true upper bound $u$ in Equation \ref{eq:Markov}, then rejecting the null hypothesis only when $u \leq \alpha$ would lead to a conservative procedure. That is, the actual false positive rate would be less than or equal to the nominal significance level $\alpha$. Using an estimate $\hat{u}$ rather than the exact value introduces the possibility that the procedure is no longer conservative and that the false positive rate is no longer controlled at the nominal level. However, from the results for the Upper Bound Method (UB) in Table~\ref{tab:sim-results}, we see that in these simulations the estimated upper bound procedure is conservative: the false positive rate is never greater than $\alpha$ under any of the settings considered. The Direct P-value Method using $\hat{p}$ (DP1) is also conservative. The Direct P-value Method based on $\hat{p}_{\text{naive}}$ (DP2) is not conservative, and the actual false positive rate exceeds the nominal level in all cases. 

Because of the conservativeness of the Upper Bound Method, it suffers from low power when the simulated jet is weak. In these cases, the Direct P-value Methods have a much higher chance of correctly rejecting the null hypothesis. However, $\hat{p} \geq 1/(1+M)$, so rejection is only possible using $\hat{p}$ if $\alpha \geq 1/(1+M)$; the power drops to zero when $\alpha$ is smaller than $1/(1+M)$, which equals $1/51$ in our simulation. Using $\hat{p}_{\text{naive}}$ allows us to reject at much lower nominal levels $\alpha$ and thus achieve reasonable power even when $\alpha$ is very small, but without properly controlling the actual false positive rate. When the jet is strong enough, the Upper Bound Method dominates the Direct P-value Methods in that it is able to achieve high power of detection at high confidence levels (low $\alpha$), while conservatively controlling the false positive rate. 

\begin{table}[p]\label{tab:sim-results}
    \caption{False positive rate and power estimated from $1000$ images simulated under the null and $1000$ images simulated under the alternative for each jet strength. Jet strengths are total expected counts in the simulated jet. The null hypothesis was rejected if $\hat{u} \leq \alpha$ for the Upper Bound Method (UB); if $\hat{p} \leq \alpha$ for the first Direct P-value Method (DP1); and if $\hat{p}_{\text{naive}} \leq \alpha$ for the second Direct P-value Method (DP2). False positive rates for DP1 and DP2 were calculated analytically. Because the null distribution was constructed from $50$ images simulated under the null, for DP1 $\hat{p} \geq 1/51$. Thus, if $\alpha < 1/51$, it is impossible for the DP1 approach to reject the null hypothesis; these cases are identified with asterisks. Boldface indicates the methods that lead to the highest power (within Monte Carlo uncertainty) while ensuring that the false positive rate is less than the nominal significance level $\alpha$. 
    }
    \begin{center}
        \begin{tabular}{cccrrrrrrr}
        \hline
        &&&\multicolumn{4}{c}{False positive rate (\%)} &
        \multicolumn{3}{c}{Power (\%)} \\
        Jet Strength & $\gamma$ (\%) & $\alpha$ (\%)& \quad UB & DP1 & DP2 && UB & DP1 & DP2 \\
        \hline
\multirow{6}{*}{20} &	 1.0 &	 2.0 &	 0.1 &	 2.0 &	 3.9 &&	 29.8 &	 \textbf{74.0} &	 81.7\\ \cline{2-10}
 &	 \multirow{2}{*}{0.5} &	 2.0 &	 0.4 &	 2.0 &	 3.9 &&	 41.4 &	 \textbf{70.9} &	 80.7\\
 &	  &	 1.0 &	 0.0 &	 0$^*$ &	 2.0 &&	 \textbf{18.1} &	 0$^*$ &	 72.0\\ \cline{2-10}
 &	 \multirow{3}{*}{0.1} &	 2.0 &	 0.8 &	 2.0 &	 3.9 &&	 48.8 &	 \textbf{67.8} &	 78.6\\
 &	  &	 1.0 &	 0.5 &	 0$^*$ &	 2.0 &&	 \textbf{33.2} &	 0$^*$ &	 66.8\\
 &	  &	 0.5 &	 0.0 &	 0$^*$ &	 2.0 &&	 \textbf{19.3} &	 0$^*$ &	 66.8\\  \hline
\multirow{6}{*}{40} &	 1.0 &	 2.0 &	 0.1 &	 2.0 &	 3.9 &&	 \textbf{99.7} &	 \textbf{100.0} &	 100.0\\ \cline{2-10}
 &	 \multirow{2}{*}{0.5} &	 2.0 &	 0.3 &	 2.0 &	 3.9 &&	 \textbf{99.7} &	 \textbf{100.0} &	 100.0\\
 &	  &	 1.0 &	 0.1 &	 0$^*$ &	 2.0 &&	 \textbf{97.6} &	 0$^*$ &	 100.0\\ \cline{2-10}
 &	 \multirow{3}{*}{0.1} &	 2.0 &	 0.8 &	 2.0 &	 3.9 &&	 \textbf{99.6} &	 \textbf{99.8} &	 100.0\\
 &	  &	 1.0 &	 0.2 &	 0$^*$ &	 2.0 &&	 \textbf{98.4} &	 0$^*$ &	 99.8\\
 &	  &	 0.5 &	 0.1 &	 0$^*$ &	 2.0 &&	 \textbf{96.2} &	 0$^*$ &	 100.0\\ \hline
\multirow{6}{*}{70} &	 1.0 &	 2.0 &	 0.1 &	 2.0 &	 3.9 &&	 \textbf{100.0} &	 \textbf{100.0} &	 100.0\\ \cline{2-10}
 &	 \multirow{2}{*}{0.5} &	 2.0 &	 0.4 &	 2.0 &	 3.9 &&	 \textbf{100.0} &	 \textbf{100.0} &	 100.0\\
 &	  &	 1.0 &	 0.0 &	 0$^*$ &	 2.0 &&	 \textbf{100.0} &	 0$^*$ &	 100.0\\ \cline{2-10}
 &	 \multirow{3}{*}{0.1} &	 2.0 &	 0.6 &	 2.0 &	 3.9 &&	 \textbf{100.0} &	 \textbf{100.0} &	 100.0\\
 &	  &	 1.0 &	 0.3 &	 0$^*$ &	 2.0 &&	 \textbf{100.0} &	 0$^*$ &	 100.0\\
 &	  &	 0.5 &	 0.2 &	 0$^*$ &	 2.0 && 	 \textbf{99.9} &	 0$^*$ &	 100.0\\
        \hline
        \end{tabular}
    \end{center}
\end{table}

\subsection{Data analysis}\label{sub:data}

\begin{figure}[t]
% \vspace{-.4cm}
\centering

\begin{minipage}[b]{0.45\linewidth}
\centering
(a)

\includegraphics[width=\textwidth]{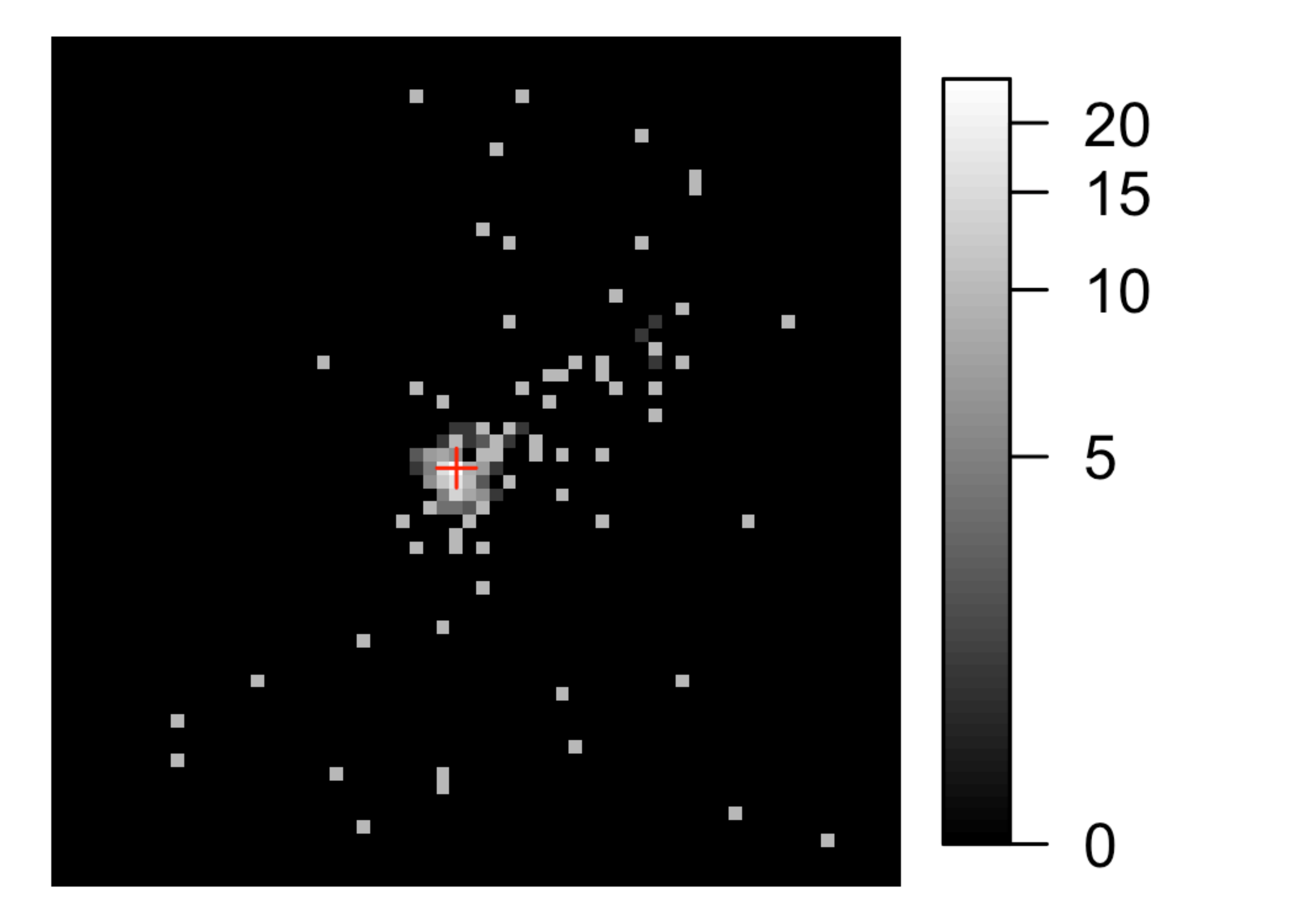}
\end{minipage}
\begin{minipage}[b]{0.45\linewidth}
\centering
(b)

\includegraphics[width=\textwidth]{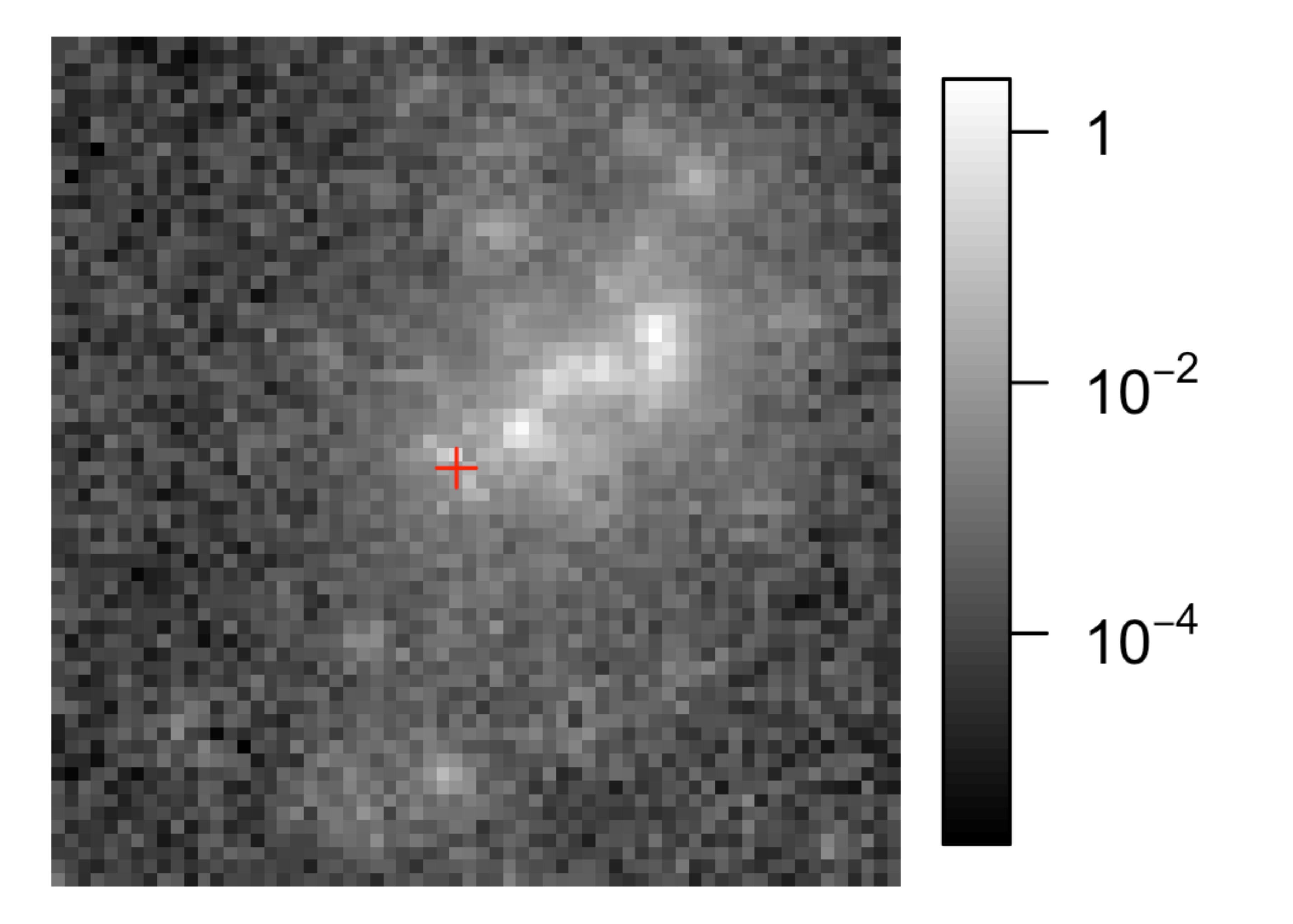}
\end{minipage}

\caption{(a) X-ray observation of the 0730+257 quasar and its possible jet. The image is centered on the location of the quasar, and its width and height are both $15.7$ arcsec. (b) Pixel-wise posterior means of the added component in the alternative model. The red + in (a) and (b) identifies the location of the quasar obtained from fitting the null model in Sherpa.}
\label{fig:obs_quasar}
\end{figure}

The X-ray jet associated with the 0730+257 quasar (redshift $z=2.868$) was observed by {\it Chandra} (ACIS-S detector) on 2009 December 12 (ObsID 10307) for about 20~ksec. We reprocessed the {\it Chandra} data in CIAO \citep{fruscione:etal:06} using the calibration database CALDB 4.5.7,\footnote{\tt{ http://cxc.cfa.harvard.edu/caldb/}} binned the original event files, selecting only the events in the energy range of 0.5--7~keV, and created a $64 \times 64$ pixel image with pixel size of 0.246~arcsec centered on the quasar; see Fig.~\ref{fig:obs_quasar}(a). The PSF, shown in Figure~\ref{fig:images}(c), was binned to the same scale. The baseline component used in fitting the alternative model included a Gaussian model of the quasar with standard deviation of 0.5 and 225 expected counts and a uniform background with 44 expected counts. The baseline component and simulated null images were created using Sherpa \citep{freeman:etal:01}.

Figure~\ref{fig:obs_quasar}(b) shows the posterior means of $\tau_1 \Lambda_{1j}$. There appears to be additional structure beyond the baseline component. The posterior distributions, $\pi(\xi \mid \by\obs)$ and $\pi(\xi \mid \by_0^{(j)})$, are shown in the bottom panel of Fig.~\ref{fig:posteriors}; $\pi(\xi \mid \by\obs)$ appears slightly more extreme relative to $g(\xi)$ than does the corresponding posterior distribution for the simulated medium jet in the middle panel of Fig.~\ref{fig:posteriors}. From Fig.~\ref{fig:upper_bounds}, it appears that the strength of evidence for additional emission beyond $\bLam_0$ is between that in the simulations with the medium jet and with the strong jet.

\begin{figure}[t]
% \vspace{-.4cm}
\centering

\begin{minipage}[b]{0.45\linewidth}
\centering
\hskip1in(a)

\kern -15pt
\includegraphics[width=\textwidth]{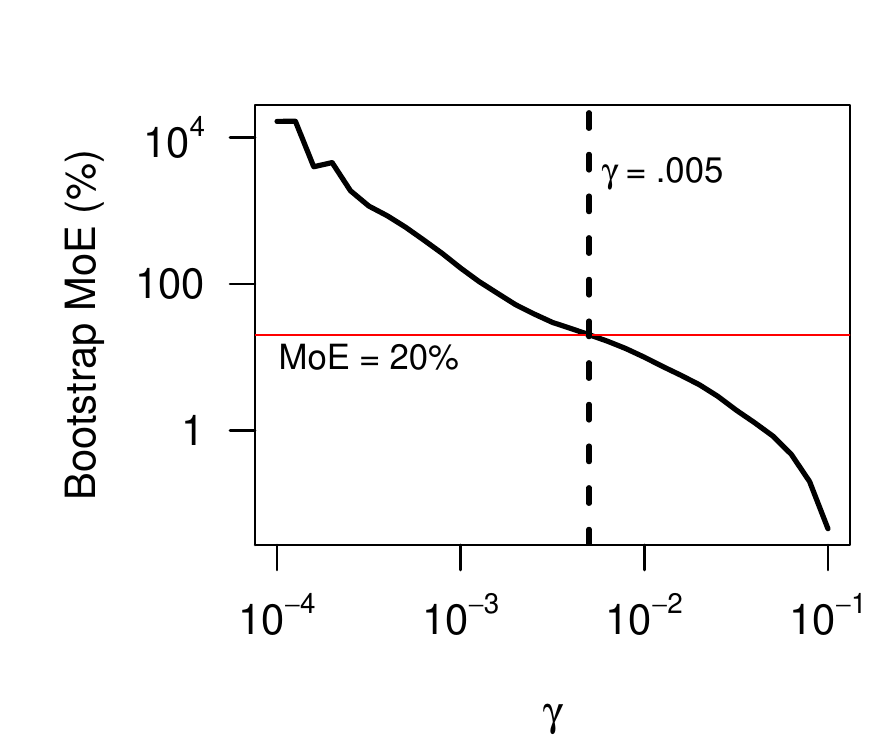}
\end{minipage}
\begin{minipage}[b]{0.45\linewidth}
\centering
\hskip1in(b)

\kern -15pt
\includegraphics[width=\textwidth]{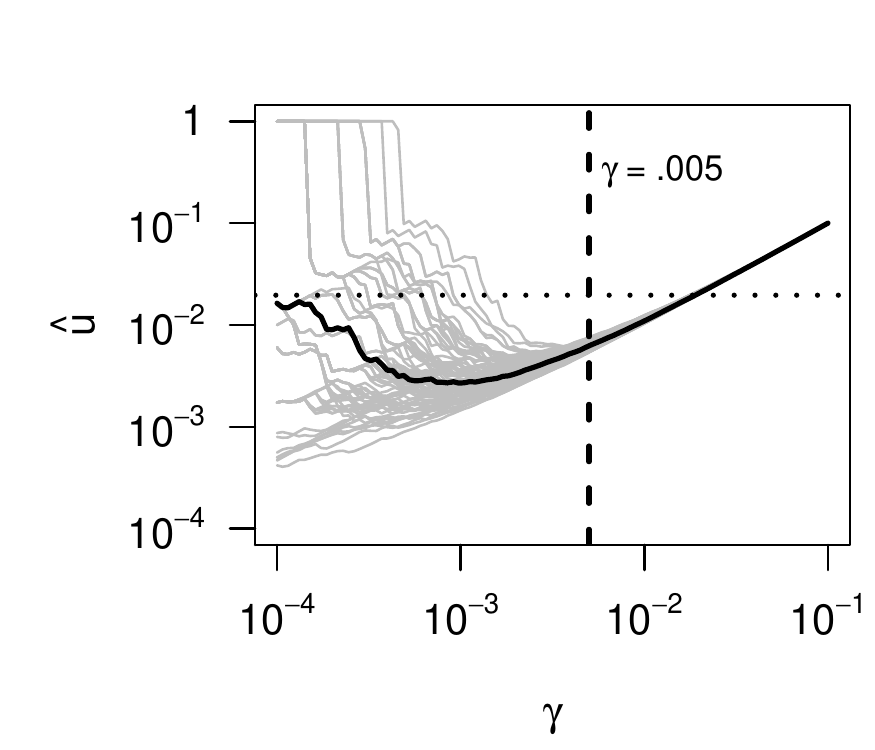}
\end{minipage}

\caption{(a) Bootstrap margins of error (MoE) of the estimated upper bound $\hat{u}$ for different values of $\gamma$ (solid black line). (b) The estimated upper bound $\hat{u}$  (solid line) and for 100 bootstrap replications (gray solid lines), obtained by sampling with replacement from the original null replicate images. The dotted horizontal line is at the minimum obtainable direct Monte Carlo p-value with $M=50$, $\hat{p}=1/51$.
The dashed vertical line in (a) and (b) at $\gamma = 0.005$ corresponds to a bootstrap MoE of 20\% (red horizontal line in (a)). 
}
\label{fig:bootstrap}
\end{figure}

Here, we perform a global test of adequacy of the null hypothesis, based on $\xi$ given in Equation~\ref{eq:fraction_intensity}, computed for the entire image. In \citet{mcke:etal:15}, we perform region-specific tests for this same observed image, based on $\xi_R$ given in Equation~\ref{eq:fraction_intensity_local}, for several regions $R$ chosen with guidance from radio observations.

We use bootstrap resampling to investigate the uncertainty in the estimated upper bound $\hat{u}$. In particular, we resampled  the $\{\by_0^{(j)}, j=1,\ldots M\}$ and their corresponding posterior samples, $\{\xi^{(j,1)}, \ldots, \xi^{(j,L)}\}$, to obtain 1000 bootstrap replications of our sample of size $LM$ from $g(\xi)$.  This bootstrap procedure treats the posterior samples, $\{\xi^{(j,1)}, \ldots, \xi^{(j,L)}\}$, given each null dataset $\by_0^{(j)}$, as fixed and estimates the uncertainties in $\hat{u}$ due to only having $M=50$ null datasets.  Figure~\ref{fig:bootstrap}(a) shows the margins of error for approximate 95\% bootstrap confidence intervals, $\exp\{\ln(\hat u) \pm 2s\}$, where $s$ is the bootstrap standard deviation of $\ln(\hat u)$. As expected, the uncertainty in $\hat{u}$ grows as $\gamma$ decreases. This is also reflected in Figure~\ref{fig:bootstrap}(b), which shows that the variability in $\hat{u}$ renders it unreliable for very small values of $\gamma$. 

Based on the estimated margins of error (MoE) in Figure~\ref{fig:bootstrap}(a), we choose to set $\gamma = 0.005$, since it is a small value that nonetheless allows a relatively precise estimate of $\hat u$, with MoE $ = 20\%$ and a 95\% bootstrap confidence interval of $(\hat u/1.2, 1.2 \hat u)$. This leads to an estimated threshold of $\hat{c}=0.073$ and an estimated upper bound on the p-value of $\hat{u}=0.0062$. For comparison, using $\gamma = 0.005$, the estimated upper bounds for the weak, medium, and strong jet simulations of Section~\ref{sub:three-images} were $0.1837$, $0.0076$, and $0.00501$, respectively. Thus, this analysis suggests significant inadequacy in the null hypothesis for the quasar, lending plausibility to the claim of a jet.

\section{Discussion} \label{sec:disc}

We have presented a method for computing the statistical significance of departures from a null model of an image. The test statistic is based on the posterior distribution under a Bayesian model that accounts for a PSF, detector inefficiencies, and Poisson noise, making it appropriate for low-count images in high-energy astrophysics. The Bayesian model allows for flexible departures from the null model via an added multiscale component. Because we use a posterior tail probability as a test statistic, we can compute an upper bound on a p-value that enables achieving high significance levels even when we have limited resources to devote to computations under the null hypothesis. We apply this method to an observed image of the 0730+257 quasar and find significant evidence of additional structure beyond the quasar and (flat) background, supporting the claim of an X-ray jet.

The simulations in Section~\ref{sub:power} illustrate the trade-off between statistical efficiency and computational efficiency that our proposed Upper Bound Method navigates. The Upper Bound Method sacrifices some statistical efficiency, as seen in the reduced power relative to the Direct P-value Methods when there is a weak jet (the first, second, and fourth rows of Table~\ref{tab:sim-results}). In return, the Upper Bound Method gains computational efficiency by enabling us to draw stronger conclusions when there are constraints on the number $M$ of simulated null images that we can afford to analyze. In particular, the Upper Bound Method enables testing at significance levels $\alpha$ smaller than $1/(1+M)$ (which the Direct P-value Method based on $\hat{p}$ cannot do), while ensuring that the false positive rate is no larger than the nominal significance level $\alpha$ (which the Direct P-value Method based on $\hat{p}_{\text{naive}}$ cannot do). Put another way, {\it for a fixed computational time, the Upper Bound Method allows valid testing at smaller significance levels than does the Direct P-value Method.}

Because of this sacrifice in statistical power, the Upper Bound Method may be too conservative to recommend if the goal is detection of very weak signals. Of course, if the signal is actually weak, the p-value under any test will most likely not be very small and can be computed with a smaller number of null simulations, so a direct Monte Carlo $\hat{p}$ may not be too computationally expensive. The real advantage of the upper bound approach is that high significance levels can be achieved for moderate or strong signals without extreme demands for simulation under the null, but with control of false positive rates.

We emphasize that it is only advantageous to use the Upper Bound Method instead of the Direct P-value Method when, due to computational constraints, we can only afford a modest $M$. The direct Monte Carlo estimates $\hat{p}$ and $\hat{p}_{\text{naive}}$ converge to the correct p-value as $M \rightarrow \infty$, while the estimated upper bound $\hat{u}$ converges to a conservative bound. In a constrained setting, however, the behavior as $M \rightarrow \infty$ is less relevant than the behavior for small $M$. When $M$ is small, $\hat{p}$ may be more conservative than $\hat{u}$ (e.g., the entries in Table~\ref{tab:sim-results} in which the false positive rate and power of the Direct P-value Method based on $\hat{p}$ are exactly zero). 

\begin{figure}[p]
\centering
\includegraphics[width=.7\linewidth]{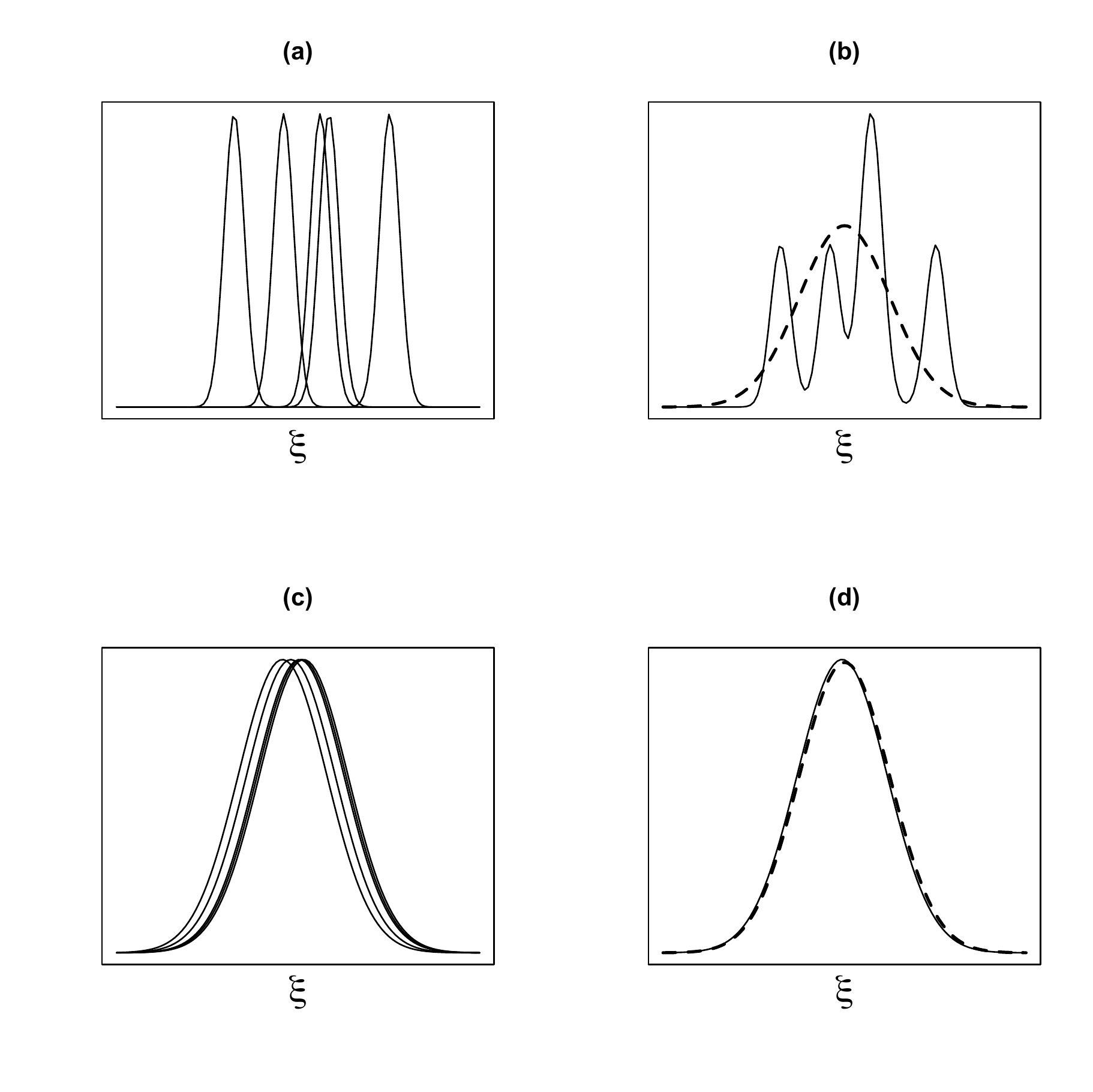}
\caption{Illustration of the possibility of obtaining an accurate estimate of a small upper bound $\hat{u}$ with small $M$. Panels (a) and (c) display posterior densities $\pi(\xi \mid \by_0^{(j)})$ for $j=1,\ldots, 5=M$, where  $\by_0^{(j)}$ are replicate images. Panels (b) and (d) show the estimated $\hat{g}(\xi) = M^{-1} \sum_{j=1}^M \pi(\xi \mid \by_0^{(j)})$ (solid lines) using the samples from Panels (a) and (c), respectively, along with the true $g(\xi)$ (dashed lines). In (a) and (b), 
there is more variation between posterior distributions of $\xi$ for different replicate images $\by_0^{(j)}$ than within the posterior of $\xi$ for any given $\by_0^{(j)}$,
and the resulting estimate of $g(\xi)$ is not close to the truth with such a small $M$; in this case, we cannot accurately estimate quantiles of $g(\xi)$. In (c) and (d), the reverse is true, and even with only $M=5$ replicate images, we can accurately estimate quantiles of $g(\xi)$.
}
\label{fig:between-within}
\end{figure}

The advantages of the Upper Bound Method relative to the Direct P-value Method depend on the reliability of the estimate $\hat{c}$. The Monte Carlo estimate $\hat{p}$  is based on  $M$ draws of the test statistic $\widehat{T}_c(\by_0^{(j)})$ given a fixed $c$. The $L$ draws from the posterior distribution conditional on $\by_0^{(j)}$ are only used to compute $\widehat{T}_c(\by_0^{(j)})$. In contrast, the upper bound $\hat{u}$ requires an estimate, $\hat{c}$, of a quantile of $g(\xi)$, and the $LM$ posterior draws given the $M$ simulated replicate images can be viewed as a cluster sample from $g(\xi)$. 
Whereas a large value of $M$ is always needed to obtain a small $\hat p$, it may be possible to obtain an accurate estimate of a small upper bound with a relatively small $M$. (Technically, this requires the posterior variance of $\xi$ for any given $\by_0$ to dominate the variability of the posterior expectation of $\xi$ as a function of $\by_0$; see Fig.~\ref{fig:between-within}.)

There are many avenues for future work. We are especially interested in exploring extensions of this method to automatically identify localized regions of significant departures from the null. This is important in astronomy because of the prevalence of low-count images and the great temptation to (over) interpret features in smoothed images (whether smoothed by eye or by an algorithm) as newly discovered objects. We are also interested in effects of model misspecification, in particular of the PSF, which may have non-negligible uncertainty. Finally, we seek to extend this type of analysis to cases where some property defining the source (e.g., the prevalence of substructure in a solar coronal loop) is tested, not just its intensity.

\subsubsection*{Acknowledgments}

This research has made use of data obtained from the {\it Chandra} Data Archive and software provided by the {\it Chandra} X-ray Center (CXC) in the application packages CIAO and Sherpa.
This work was conducted under the auspices of the CHASC International Astrostatistics Center. CHASC is supported by NSF grants DMS 1208791, DMS 1209232, DMS 1513492, DMS 1513484, and DMS 1513546. We acknowledge support from SI's Competitive Grants Fund 40488100HH0043. DvD acknowledges support from a Wolfson Research Merit Award provided by the British Royal Society and from a Marie-Curie Career Integration Grant provided by the European Commission, and VK and AS  from a NASA contract to the {\it Chandra} X-Ray Center NAS8-03060. In addition, we thank CHASC members for many helpful discussions, especially Xiao-Li Meng and Kathryn McKeough.

\appendix
\section{Technical details: Computing $u$}\label{sec:technical}

In this appendix, we derive the p-value upper bounds of Equations~\ref{eq:Markov} and \ref{eq:Markovppp}. We begin with the case in which the null hypothesis contains no unknown parameters. Because $T_c(\by)$ is non-negative, Markov's inequality\footnote{Markov's inequality states that if $X$ is a non-negative random variable and $a > 0$, then $\Pr(X \geq a) \leq E(X)/a$.} yields
\begin{equation}
      p \leq \frac{E\l\{T_c(\by_0) \r\}}{T_c(\by\obs)} = u,
\end{equation}
where the expectation $E\{T_c(\by_0)\} = \sum_{\by_0} T_c(\by_0)\, \like_0(\by_0)$. Using the definition of $T_c$, we can rewrite this expectation as
\begin{align}
    E\l\{T_c(\by_0) \r\} 
    % &= E\l\{\Pr\l(\xi \geq c \mid \by_0\r) \r\} \nonumber \\
    &= \sum_{\by_0} \Pr(\xi \geq c \mid \by_0) \, \like_0(\by_0) \nonumber \\
    &= \sum_{\by_0} \l\{\int_c^1 \pi(\xi \mid \by_0) \, {\rm d}\xi \r\}\, \like_0(\by_0) \nonumber \\
    &= \int_c^1 \l\{\sum_{\by_0}  \pi(\xi \mid \by_0) \, \like_0(\by_0) \r\} \, {\rm d}\xi  \nonumber \\
    &= \Pr(\xi \geq c), \label{eq:prob}
\end{align}
where the probability in Equation \ref{eq:prob} is taken with respect to $g(\xi)$ as given in Equation~\ref{eq:expost}.

If the null hypothesis contains unknown parameters $\btheta_0$, then we obtain Equations~\ref{eq:Markovppp} and \ref{eq:expostpred} as follows. For each fixed $\btheta_0$, Markov's inequality yields
$$
p(\btheta_0) \leq \frac{E\l\{T_c(\by_0) \mid \btheta_0\r\}}{T_c(\by\obs)},
$$
where the expectation in the numerator is taken with respect to $\like_0(\by_0 \mid \btheta_0)$. Averaging over $\pi_0(\btheta_0 \mid \by\obs)$, the ppp-value is bounded by
$$
\textrm{ppp-value} \leq \frac{E\l\{T_c(\by_0) \mid \by\obs\r\}}{T_c(\by\obs)} = u_{\rm ppp}.
$$
The expectation
\begin{equation}\label{eq:tail}
    E\l\{T_c(\by_0) \mid \by\obs\r\} = \Pr(\xi \geq c \mid \by\obs),
\end{equation}
where the probability is taken with respect to the distribution in Equation~\ref{eq:expostpred}.

\bigskip
\noindent

\bibliographystyle{natbib}
\bibliography{lira}

\begin{thebibliography}{}

\bibitem[Bayarri and Castellanos(2007)]{bayarri:castellanos:07}
Bayarri, M.~J. and Castellanos, M.~E. (2007).
\newblock Bayesian checking of the second levels of hierarchical models (with
  discussion).
\newblock \emph{Statistical Science} \textbf{22}, 3, 322--367.

\bibitem[Berger and Delampady(1987)]{berg:dela:87}
Berger, J.~O. and Delampady, M. (1987).
\newblock Testing precise hypotheses (with discussion).
\newblock \emph{Statistical Science} \textbf{2}, 317--352.

\bibitem[{Bull} \emph{et~al.}(2014){Bull}, {Wehus}, {Eriksen}, {Ferreira},
  {Fuskeland}, {Gorski}, and {Jewell}]{bull:etal:14}
{Bull}, P., {Wehus}, I.~K., {Eriksen}, H.~K., {Ferreira}, P.~G., {Fuskeland},
  U., {Gorski}, K.~M., and {Jewell}, J.~B. (2014).
\newblock {A CMB Gibbs sampler for localized secondary anisotropies}.
\newblock \emph{arxiv:1410.2544} .

\bibitem[Calderwood \emph{et~al.}(2001)Calderwood, Dobrzycki, Jessop, and
  Harris]{cald:etal:01}
Calderwood, T., Dobrzycki, A., Jessop, H., and Harris, D.~E. (2001).
\newblock The sliding-cell detection program for {C}handra x-ray data.
\newblock In F.~R. Harnden, F.~A. Primini, and H.~E. Payne, eds.,
  \emph{Astronomical Data Analysis Software and Systems X}, vol. 103,  443.
  ASP, San Francisco.

\bibitem[Connors and van Dyk(2007)]{conn:vand:07}
Connors, A. and van Dyk, D.~A. (2007).
\newblock How to win with non-{G}aussian data: {P}oisson goodness-of-fit.
\newblock In \emph{Statistical Challenges in Modern Astronomy IV {\rm (Editors:
  E.~Feigelson and G.~Babu)}}, vol. CS371,  101--117. Astronomical Society of
  the Pacific, San Francisco.

\bibitem[Davison and Hinkley(1997)]{davison:hinkley:97}
Davison, A.~C. and Hinkley, D.~V. (1997).
\newblock \emph{Bootstrap methods and their application}.
\newblock Cambridge University Press, Cambridge, United Kingdom.

\bibitem[{Ebeling} and {Wiedenmann}(1993)]{ebel:wied:93}
{Ebeling}, H. and {Wiedenmann}, G. (1993).
\newblock {Detecting structure in two dimensions combining Voronoi tessellation
  and percolation}.
\newblock \emph{Physical Review E} \textbf{47}, 704--710.

\bibitem[Esch \emph{et~al.}(2004)Esch, Connors, Karovska, and van
  Dyk]{esch:etal:04}
Esch, D.~N., Connors, A., Karovska, M., and van Dyk, D.~A. (2004).
\newblock An image restoration technique with error estimates.
\newblock \emph{The Astrophysical Journal} \textbf{610}, 1213--1227.

\bibitem[Freeman \emph{et~al.}(2001)Freeman, Doe, and
  Siemiginowska]{freeman:etal:01}
Freeman, P., Doe, S., and Siemiginowska, A. (2001).
\newblock Sherpa: a mission-independent data analysis application.
\newblock \emph{Proceedings of the International Society for Optical
  Engineering} \textbf{4477}, 76--87.

\bibitem[{Freeman} \emph{et~al.}(2002){Freeman}, {Kashyap}, {Rosner}, and
  {Lamb}]{free:etal:02}
{Freeman}, P.~E., {Kashyap}, V., {Rosner}, R., and {Lamb}, D.~Q. (2002).
\newblock A wavelet-based algorithm for the spatial analysis of {P}oisson data.
\newblock \emph{Astrophysical Journal Supplement Series} \textbf{138},
  185--218.

\bibitem[Friedenberg and Genovese(2013)]{friedenberg:genovese:13}
Friedenberg, D.~A. and Genovese, C.~R. (2013).
\newblock Straight to the source: Detecting aggregate objects in astronomical
  images with proper error control.
\newblock \emph{Journal of the American Statistical Association} \textbf{108},
  502, 456--468.

\bibitem[Friston \emph{et~al.}(1995)Friston, Holmes, Worsley, Poline, Frith,
  and Frackowiak]{friston:etal:95}
Friston, K.~J., Holmes, A.~P., Worsley, K.~J., Poline, J.-P., Frith, C.~D., and
  Frackowiak, R.~S.~J. (1995).
\newblock Statistical parametric maps in functional imaging: A general linear
  approach.
\newblock \emph{Human Brain Mapping} \textbf{2}, 189--210.

\bibitem[{Fruscione} \emph{et~al.}(2006){Fruscione}, {McDowell}, {Allen},
  {Brickhouse}, {Burke}, {Davis}, {Durham}, {Elvis}, {Galle}, {Harris},
  {Huenemoerder}, {Houck}, {Ishibashi}, {Karovska}, {Nicastro}, {Noble},
  {Nowak}, {Primini}, {Siemiginowska}, {Smith}, and {Wise}]{fruscione:etal:06}
{Fruscione}, A., {McDowell}, J.~C., {Allen}, G.~E., {Brickhouse}, N.~S.,
  {Burke}, D.~J., {Davis}, J.~E., {Durham}, N., {Elvis}, M., {Galle}, E.~C.,
  {Harris}, D.~E., {Huenemoerder}, D.~P., {Houck}, J.~C., {Ishibashi}, B.,
  {Karovska}, M., {Nicastro}, F., {Noble}, M.~S., {Nowak}, M.~A., {Primini},
  F.~A., {Siemiginowska}, A., {Smith}, R.~K., and {Wise}, M. (2006).
\newblock {CIAO: Chandra's data analysis system}.
\newblock In \emph{Society of Photo-Optical Instrumentation Engineers (SPIE)
  Conference Series}, vol. 6270 of \emph{Society of Photo-Optical
  Instrumentation Engineers (SPIE) Conference Series}, ~1.

\bibitem[Gelman \emph{et~al.}(1996)Gelman, Meng, and
  Stern]{gelman:meng:stern:96}
Gelman, A., Meng, X.-L., and Stern, H. (1996).
\newblock Posterior predictive assessment of model fitness via realized
  discrepancies (with discussion).
\newblock \emph{Statistica Sinica} \textbf{6}, 733--807.

\bibitem[Godtliebsen \emph{et~al.}(2004)Godtliebsen, Marron, and
  Chaudhuri]{godtliebsen:etal:04}
Godtliebsen, F., Marron, J.~S., and Chaudhuri, P. (2004).
\newblock Statistical significance of features in digital images.
\newblock \emph{Image and Vision Computing} \textbf{22}, 1093--1104.

\bibitem[Gross and Vitells(2010)]{gros:vite:10}
Gross, E. and Vitells, O. (2010).
\newblock Trial factors for the look elsewhere effect in high energy physics.
\newblock \emph{The European Physical Journal C} \textbf{70}, 525--530.

\bibitem[Guan(2008)]{guan:08}
Guan, Y. (2008).
\newblock A goodness-of-fit test for inhomogeneous spatial {P}oisson processes.
\newblock \emph{Biometrika} \textbf{95}, 4, 831--845.

\bibitem[Harris and Krawczynski(2006)]{harris:06}
Harris, D.~E. and Krawczynski, H. (2006).
\newblock X-ray emission from extragalactic jets.
\newblock \emph{Annual Review of Astronomy and Astrophysics} \textbf{44},
  463--506.

\bibitem[Holmstr\"{o}m and Pasanen(2012)]{holmstrom:pasanen:12}
Holmstr\"{o}m, L. and Pasanen, L. (2012).
\newblock Bayesian scale space analysis of differences in images.
\newblock \emph{Technometrics} \textbf{54}, 1, 16--29.

\bibitem[Jasche and Wandelt(2013)]{jasche:wand:13}
Jasche, J. and Wandelt, B.~D. (2013).
\newblock Methods for bayesian power spectrum inference with galaxy surveys.
\newblock \emph{The Astrophysical Journal} \textbf{779}, 1, 15.

\bibitem[Lucy(1974)]{lucy:74}
Lucy, L.~B. (1974).
\newblock An iterative technique for the rectification of observed
  distributions.
\newblock \emph{The Astronomical Journal} \textbf{79}, 745--754.

\bibitem[McKeough \emph{et~al.}(2014)McKeough, Kashyap, and
  McKillop]{mcke:poster}
McKeough, K., Kashyap, V., and McKillop, S. (2014).
\newblock Quantifying the significance of substructure in coronal loops.
\newblock Poster presented at the American Geophysical Union Fall Meeting 2015,
  San Francisco, California. Final paper number SH13C-4127.

\bibitem[McKeough \emph{et~al.}(2015)McKeough, Siemiginowska,
  \emph{et~al.}]{mcke:etal:15}
McKeough, K., Siemiginowska, A., \emph{et~al.} (2015).
\newblock Chandra {X}-ray imaging of the highest-redshift quasar jets.
\newblock \emph{Manuscript} In preparation.

\bibitem[Meng(1994)]{meng:94}
Meng, X.-L. (1994).
\newblock Posterior predictive $p$-values.
\newblock \emph{The Annals of Statistics} \textbf{22}, 3, 1142--1160.

\bibitem[Nowak and Kolaczyk(2000)]{nowak:kolaczyk:00}
Nowak, R.~D. and Kolaczyk, E.~D. (2000).
\newblock A statistical multiscale framework for {P}oisson inverse problems.
\newblock \emph{IEEE Transactions on Information Theory} \textbf{46}, 5,
  1811--1825.

\bibitem[{Pina} and {Puetter}(1992)]{pina:puet:92}
{Pina}, R.~K. and {Puetter}, R.~C. (1992).
\newblock {Incorporation of Spatial Information in Bayesian Image
  Reconstruction: The Maximum Residual Likelihood Criterion}.
\newblock \emph{Publications of the Astronomical Society of the Pacific}
  \textbf{104}, 1096.

\bibitem[Protassov \emph{et~al.}(2002)Protassov, van Dyk, Connors, Kashyap, and
  Siemiginowska]{prot:etal:02}
Protassov, R., van Dyk, D.~A., Connors, A., Kashyap, V., and Siemiginowska, A.
  (2002).
\newblock Statistics: Handle with care -- detecting multiple model components
  with the likelihood ratio test.
\newblock \emph{The Astrophysical Journal} \textbf{571}, 545--559.

\bibitem[Radke \emph{et~al.}(2005)Radke, Andra, Al-Kofahi, and
  Roysam]{radke:etal:05}
Radke, R., Andra, S., Al-Kofahi, O., and Roysam, B. (2005).
\newblock Image change detection algorithms: a systematic survey.
\newblock \emph{IEEE Transactions on Image Processing} \textbf{14}, 3,
  294--307.

\bibitem[Richardson(1972)]{rich:72}
Richardson, W.~H. (1972).
\newblock {B}ayesian-based iterative method of image restoration.
\newblock \emph{Journal of the Optical Society of {A}merica} \textbf{62},
  55--59.

\bibitem[Ripley and Sutherland(1990)]{ripley:sutherland:90}
Ripley, B.~D. and Sutherland, A.~I. (1990).
\newblock Finding spiral structures in images of galaxies.
\newblock \emph{Phil. Trans. R. Soc. Lond. A} \textbf{332}, 477--485.

\bibitem[Rubin(1984)]{rubin:84}
Rubin, D.~B. (1984).
\newblock Bayesianly justifiable and relevant frequency calculations for the
  applied statistician.
\newblock \emph{The Annals of Statistics} \textbf{12}, 4, 1151--1172.

\bibitem[Sinharay and Stern(2003)]{sinharay:stern:03}
Sinharay, S. and Stern, H.~S. (2003).
\newblock Posterior predictive checking in hierarchical models.
\newblock \emph{Journal of Statistical Planning and Inference} \textbf{111},
  209--221.

\bibitem[Sutter \emph{et~al.}(2014)Sutter, Wandelt, McEwen, Bunn, Karakci,
  Korotkov, Timbie, Tucker, and Zhang]{sutter:etal:14}
Sutter, P.~M., Wandelt, B.~D., McEwen, J.~D., Bunn, E.~F., Karakci, A.,
  Korotkov, A., Timbie, P., Tucker, G.~S., and Zhang, L. (2014).
\newblock Probabilistic image reconstruction for radio interferometers.
\newblock \emph{Monthly Notices of the Royal Astronomical Society}
  \textbf{438}, 1, 768--778.

\bibitem[Sutton and Wandelt(2006)]{sutton:wand:06}
Sutton, E.~C. and Wandelt, B.~D. (2006).
\newblock Optimal image reconstruction in radio interferometry.
\newblock \emph{The Astrophysical Journal Supplement Series} \textbf{162}, 2,
  401.

\bibitem[Thon \emph{et~al.}(2012)Thon, Rue, Skr{\o}vseth, and
  Godtliebsen]{thon:etal:12}
Thon, K., Rue, H., Skr{\o}vseth, S.~O., and Godtliebsen, F. (2012).
\newblock Bayesian multiscale analysis of images modeled as {G}aussian {M}arkov
  random fields.
\newblock \emph{Computational Statistics and Data Analysis} \textbf{56},
  49--61.

\bibitem[Urry and Padovani(1995)]{urry:95}
Urry, C.~M. and Padovani, P. (1995).
\newblock Unified schemes for radio-loud active galactic nuclei.
\newblock \emph{Publications of the Astronomical Society of the Pacific}
  \textbf{107}, 803--845.

\bibitem[van Dyk(2012)]{vand:12scma}
van Dyk, D.~A. (2012).
\newblock Discussion of ``{C}osmological bayesian model selection: {R}ecent
  advances and open challenges by {R}. {T}rotta''.
\newblock In \emph{{Statistical Challenges in Modern Astronomy V {\rm (Editors:
  E.~Feigelson and G.~Babu)}}},  141--146. Springer Verlag.

\bibitem[van Dyk(2014)]{vand:14}
van Dyk, D.~A. (2014).
\newblock The role of statistics in the discovery of a {H}iggs boson.
\newblock \emph{Annual Review of Statistics and Its Application} \textbf{1},
  41--59.

\bibitem[Weinberg(2012)]{weinberg:12}
Weinberg, M.~D. (2012).
\newblock Computing the bayes factor from a markov chain monte carlo simulation
  of the posterior distribution.
\newblock \emph{Bayesian Anal.} \textbf{7}, 3, 737--770.

\bibitem[Zhao(2014)]{zhao:14}
Zhao, S. (2014).
\newblock \emph{Causal Inference and Model Selection in Complex Settings}.
\newblock Ph.D. thesis, Department of Statistics, University of California,
  Irvine.

\end{thebibliography}

\end{document}